\documentclass[a4paper,11pt, nofootinbib]{article}
\pdfoutput=1 

\usepackage{jheppub} 

\usepackage[T1]{fontenc} 

\usepackage{ bbold }
\usepackage{bm}
\usepackage{graphicx} 
\usepackage{subcaption}
\usepackage{physics}
\usepackage{tikz} 
\usepackage{pgfplots}
\pgfplotsset{compat=newest}
\DeclareUnicodeCharacter{2212}{−}
\usepgfplotslibrary{groupplots, dateplot}
\usetikzlibrary{patterns,shapes.arrows}
\usetikzlibrary{plotmarks}
\usetikzlibrary{spy}
 
\newcommand{\legendFontSize}{\footnotesize}
\newcommand{\legendMarkerScale}{3}

\newcommand{\axisLabelFontSize}{\normalsize}
\newcommand{\tickFontSize}{\small}

\newcommand{\markerScale}{3}

\pgfdeclareplotmark{o}{%
    \pgfpathcircle{\pgfpointorigin}{0.5*\pgfplotmarksize}%
    \pgfusepathqstroke
}

\pgfdeclareplotmark{square}{%
 \pgfpathmoveto{\pgfpoint{-0.5*\pgfplotmarksize}{-0.5*\pgfplotmarksize}}%
 \pgfpathlineto{\pgfpoint{0.5\pgfplotmarksize}{-0.5*\pgfplotmarksize}}
 \pgfpathlineto{\pgfpoint{0.5\pgfplotmarksize}{0.5*\pgfplotmarksize}}
 \pgfpathlineto{\pgfpoint{-0.5\pgfplotmarksize}{0.5*\pgfplotmarksize}}
 \pgfpathclose
 \pgfusepathqstroke
}
\pgfdeclareplotmark{x}{
  \pgfpathmoveto{\pgfqpoint{-0.5\pgfplotmarksize}{-0.5\pgfplotmarksize}}%
  \pgfpathlineto{\pgfqpoint{0.5\pgfplotmarksize}{0.5\pgfplotmarksize}}%
  \pgfpathmoveto{\pgfqpoint{-0.5\pgfplotmarksize}{0.5\pgfplotmarksize}}%
  \pgfpathlineto{\pgfqpoint{0.5\pgfplotmarksize}{-0.5\pgfplotmarksize}}%
  \pgfusepathqstroke
}

\pgfdeclareplotmark{Custom_8}{%
 \pgfpathmoveto{\pgfpoint{0}{-0.5*\pgfplotmarksize}}%
 \pgfpathlineto{\pgfpoint{-0.5\pgfplotmarksize}{0}}
 \pgfpathlineto{\pgfpoint{0}{0.5*\pgfplotmarksize}}
 \pgfpathclose
 \pgfusepathqstroke
} 
  
\pgfdeclareplotmark{Custom_9}{%
 \pgfpathmoveto{\pgfpoint{0}{-0.5*\pgfplotmarksize}}%
 \pgfpathlineto{\pgfpoint{0.5\pgfplotmarksize}{0}}
 \pgfpathlineto{\pgfpoint{0}{0.5*\pgfplotmarksize}}
 \pgfpathclose 
 \pgfusepathqstroke
}

\pgfdeclareplotmark{Custom_10}{%
 \pgfpathmoveto{\pgfpoint{-0.5*\pgfplotmarksize}{0}}%
 \pgfpathlineto{\pgfpoint{0.5\pgfplotmarksize}{0}}
 \pgfpathlineto{\pgfpoint{0}{0.5*\pgfplotmarksize}}
 \pgfpathclose
 \pgfusepathqstroke
}
 
\pgfdeclareplotmark{Custom_11}{%
 \pgfpathmoveto{\pgfpoint{-0.5*\pgfplotmarksize}{0}}%
 \pgfpathlineto{\pgfpoint{0.5\pgfplotmarksize}{0}}
 \pgfpathlineto{\pgfpoint{0}{-0.5*\pgfplotmarksize}}
 \pgfpathclose   
 \pgfusepathqstroke
}

\pgfplotsset{   every axis/.append style={
                                            clip marker paths=true,
                                            legend style={font=\legendFontSize,  mark options={scale=\legendMarkerScale, line join = round, line width = 1}, fill=none},
                                            xlabel style={font=\axisLabelFontSize},
                                            ylabel absolute,
                                            ylabel style={yshift=0.7em, font=\axisLabelFontSize},
                                            yticklabel style={font=\tickFontSize, /pgf/number format/fixed, /pgf/number format/precision=2, /pgf/number format/fixed zerofill},
                                            xticklabel style={font=\tickFontSize}
                                        }}
                                        
\tikzset{ every mark/.append style={ scale = \markerScale, line width=1pt, line join=round}}

\usetikzlibrary{positioning}
\usetikzlibrary{shapes.geometric}
\tikzstyle{tensorA}=[rectangle,draw=red!50,fill=red!20,thick]
\tikzstyle{tensorB}=[rectangle,draw=blue!50,fill=blue!20,thick]
\tikzstyle{tensorE}=[rectangle,draw=black!50,fill=black!20,thick]
\tikzstyle{transferA}=[rectangle,
                      draw=red!50,fill=red!20,
                      thick,
                      minimum width=4em]
\tikzstyle{transferB}=[rectangle,
                      draw=blue!50,fill=blue!20,
                      thick,
                      minimum width=4em]
\tikzstyle{transferE}=[rectangle,
                      draw=black!50,fill=black!20,
                      thick, minimum width=4em]
\tikzstyle{utensorA}=[isosceles triangle,
                      isosceles triangle apex angle=120,
                      shape border rotate=-90,
                      draw=red!50,fill=red!20,
                      thick,
                      minimum width=4em]
\tikzstyle{vtensorA}=[isosceles triangle,
                      isosceles triangle apex angle=120,
                      shape border rotate=90,
                      draw=red!50,fill=red!20,
                      thick,
                      minimum width=4em]
\tikzstyle{utensorB}=[isosceles triangle,
                      isosceles triangle apex angle=120,
                      shape border rotate=-90,
                      draw=blue!50,fill=blue!20,
                      thick,
                      minimum width=4em]
\tikzstyle{vtensorB}=[isosceles triangle,
                      isosceles triangle apex angle=120,
                      shape border rotate=90,
                      draw=blue!50,fill=blue!20,
                      thick,
                      minimum width=4em]

\title{Entanglement entropy and negativity in the Ising model with defects }

\author[a,1]{David Rogerson\note{Corresponding author.}}
\author[a,b]{Frank Pollmann}
\author[c,2]{Ananda Roy\note{Corresponding author.}}

\affiliation[a]{Department of Physics, T42, Technische Universit\"at M\"unchen,\\ 85748 Garching, Germany}
\affiliation[b]{Munich Center for Quantum Science and Technology (MCQST),\\ 80799 Munich, Germany}
\affiliation[c]{Department of Physics and Astronomy, Rutgers University,\\ Piscataway, NJ 08854-8019 USA}

\emailAdd{david.rogerson@tum.de}
\emailAdd{frank.pollmann@tum.de}
\emailAdd{ananda.roy@physics.rutgers.edu}

\abstract{Defects in two-dimensional conformal field theories~(CFTs) contain signatures of their characteristics. In this work, we analyze entanglement properties of subsystems in the presence of energy and duality defects in the Ising CFT using the density matrix renormalization group~(DMRG) technique. In particular, we compute the entanglement entropy~(EE) and the entanglement negativity~(EN) in the presence of defects. For the EE, we consider the cases when the defect lies within the subsystem and at the edge of the subsystem. We show that the EE for the duality defect exhibits fundamentally different characteristics compared to the energy defect due to the existence of localized and delocalized zero energy modes. Of special interest is the nontrivial `finite-size correction' in the EE obtained recently using free fermion computations~\cite{Roy2021a}. These corrections arise when the subsystem size is appreciable compared to the total system size and lead to a deviation from the usual logarithmic scaling characteristic of one-dimensional quantum-critical systems. Using matrix product states with open and infinite boundary conditions, we numerically demonstrate the disappearance of the zero mode contribution for finite subsystem sizes in the thermodynamic limit. Our results provide further support to the recent free fermion computations, but clearly contradict earlier analytical field theory calculations based on twisted torus partition functions. Subsequently, we compute the logarithm of the EN~(log-EN) between two disjoint subsystems separated by a defect. We show that the log-EN scales logarithmically with the separation of the subsystems. However, the coefficient of this logarithmic scaling yields a continuously-varying effective central charge that is different from that obtained from analogous computations of the EE. The defects leave their fingerprints in the subleading term of the scaling of the log-EN. Furthermore, the log-EN receives similar `finite size corrections' like the EE which leads to deviations from its characteristic logarithmic scaling.}

\begin{document} 
\maketitle
\flushbottom

\section{Introduction}
\label{sec:intro}
Entanglement plays a central role in the development of long-range correlations in quantum critical phenomena. Thus, quantification of the entanglement in a quantum-critical system provides a way to characterize the universal properties of the critical point. For 1+1D quantum-critical systems described by conformal field theories~(CFTs), entanglement measures provide a viable way to extract characteristics of the CFT. For instance, the von-Neumann entropy~($\mathcal{S}_A$) in the ground state~[{\it i.e.}, the entanglement entropy~(EE)] for a subsystem~(A) exhibits universal logarithmic scaling with the subsystem size~\cite{Holzhey1994, Calabrese2004}. Here, 
\begin{equation}
\label{EE_def}
\mathcal{S}_A = -{\rm Tr}\rho_A\ln\rho_A,\ \rho_A = {\rm Tr}_B\rho,
\end{equation}
where the total system with density matrix $\rho$ is partitioned into A and B. The coefficient of this scaling determines a fundamental  property of the CFT:  the central charge, which quantifies, crudely speaking, the number of long-wavelength degrees of freedom. The aforementioned scaling, together with strong subaddivity property of entropy~\cite{Casini2004} and Lorentz invariance, leads to an alternate proof~\cite{Casini2006} of the celebrated $c$-theorem~\cite{Zamolodchikov1986} in 1+1 dimensions. At the same time, the scaling of the EE in these gapless systems~\cite{Calabrese2008, Tagliacozzo2008, Pollmann2009} and their gapped counterparts~\cite{Hastings2007} lies at the heart of the success of numerical techniques like density matrix renormalization group~(DMRG)~\cite{White1992, Schollwock2011} in simulating 1+1D quantum systems. 

Given the widespread success of the EE in characterizing bulk properties of quantum-critical points, it is natural to ask if the EE also captures signatures of boundaries in conformal-invariant systems~\cite{Roy2021b}. For finite systems with boundaries at a conformal critical point, the EE receives a universal, subleading, boundary-dependent contribution, the so-called `boundary entropy'~\cite{Affleck1991,Calabrese2009}. The latter, related to the `ground-state degeneracy' of the system, plays a central role in a wide-range of problems both in condensed matter physics~\cite{Affleck1995, Saleur1998} and in string theory~\cite{Gaberdiel2000}. The boundary contribution in the EE is a valuable diagnostic for identifying the different boundary fixed points of a given  CFT~\cite{Affleck2009, Cardy2016, Roy2020a}. In particular, the Ishibashi/Cardy states for the rational CFTs~\cite{Cardy1989} lead to a direct computation of the different boundary entropies~\cite{Cardy2016, Roy2021b}. 

Defects constitute the more general setting where one CFT, instead of being terminated with a given boundary condition, is glued to another CFT. Despite being ubiquitous, such defects are relatively less understood as far as their entanglement properties are concerned. A major difficulty is that lattice Hamiltonians for defects of rational CFTs are not yet available, although some progress has been made in this direction~\cite{Belletete2018, Belletete2020, Aasen2020}. The notable exception is the free, real fermion~(Ising) CFT, for which the defects have been completely classified and corresponding lattice Hamiltonians have been found~\cite{Henkel1989, Baake1989, Grimm1990, Oshikawa1997, Grimm2001}. Of particular interest are topological or purely transmissive defects. Such defects commute with the generators of conformal transformations~\cite{Petkova2000, Bachas2001, Frohlich2004, Frohlich2006,Aasen2016} and thus, can be deformed without affecting the values of the correlation functions as long as they are not taken across field insertions~(hence the moniker topological). They reflect the internal symmetries of the CFT and relate the order-disorder dualities of the CFT to the high-low temperature dualities of the corresponding off-critical model~\cite{Frohlich2004, Krammers1941,Savit1980}. They also play an important role in the study of anyonic chains and in the correspondence between CFTs and three-dimensional topological field theories~\cite{Buican2017}. 
\begin{figure}
    \centering
    \includegraphics[width = \textwidth]{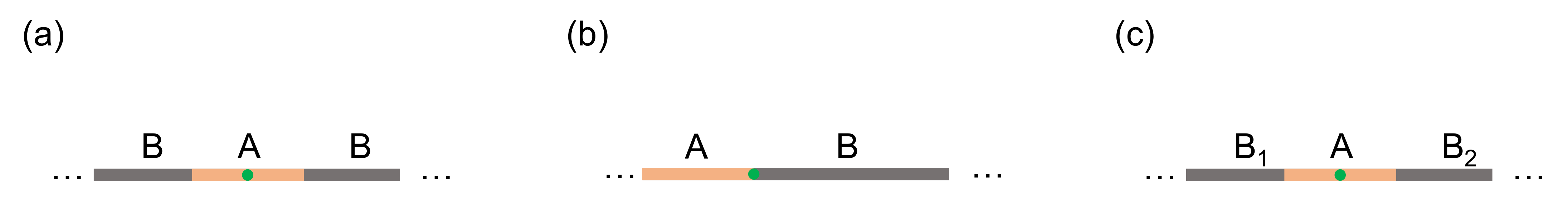}
    \caption{\label{fig:geoms} Geometric arrangements of the subsystem(s) and the rest of the system considered for the bipartite EE~(a,b) and the log-EN~(c) computations. The defect is indicated by a green dot. (a) Subsystem A is located symmetrically around the defect. (b) The defect lies at the interface between the subsystem~(A) and the rest~(B). (c) Two subsystems, $B_1$ and $B_2$, separated by a segment A, with the defect symmetrically located within A.}
\end{figure}

The importance of defects in the analysis of CFTs provides a natural motivation for the investigation of their entanglement properties. Consider the EE of a subsystem for two distinct geometries~[see Fig.~\ref{fig:geoms}(a,b)]: i) where the defect is symmetrically located within the subsystem and ii) when the defect is located at the interface between the subsystem and the rest. For case i), as long as the defect sufficiently far from the edge of the subsystem, the leading order dependence of the EE~(referred to as symmetric EE), is governed entirely by the bulk correlations around the edge of the subsystem. The subleading term depends on the defect. After usual folding maneuvers, the subleading term can be equated to a boundary entropy with double the bulk degrees of freedom~\cite{Oshikawa1997, Saleur1998, Saleur2000}. This leads to predictions for the subleading term in the symmetric EE for defects in rational CFTs~\cite{Gutperle2015} when the size of the subsystem is much smaller than the system size. The situation for case ii) is more subtle. In particular, both the leading and subleading order terms of the EE~(referred to as interface EE) depend on the defect. The leading order term has been computed using the replica trick for the free, compactified boson CFT~\cite{Sakai2008} and the Ising CFT~\cite{Brehm2015}~(see also~\cite{Eisler2010, Peschel2012e, Calabrese2011ru}). The coefficient of this leading order term defines an effective central charge that depends continuously on the defect strength. The subleading term, initially computed only for the topological defects~\cite{Gutperle2015, Brehm2015} using the corresponding twisted torus partition functions~\cite{Petkova2000}, was shown recently to be incorrect in Ref.~\cite{Roy2021a} for the Ising CFT using free-fermion computations. 

In this work, we analyze the entanglement properties of subsystems for the Ising CFT~\cite{Roy2021a} with energy and duality defects using the density matrix renormalization group~(DMRG) technique. The primary goals of this work are the following. First, DMRG provides a completely independent check of the free-fermion computations performed for the topological defect~(a special case of the duality defect) in Ref.~\cite{Roy2021a}. Second, DMRG allows numerical computations to be performed in both finite and infinite systems by appropriately choosing boundary conditions~(bcs)~\cite{Schollwock2011,Hauschild2018}. As shown in Ref.~\cite{Roy2021a}, the existence of a nonlocal zero mode in the topological case leads to nontrivial `finite size corrections'~\cite{Klich2017} for both the symmetric and interface EEs. Here, we show that the same is also true for general duality defects away from the topological point~\cite{Roy2021b}. Furthermore, by choosing infinite bc in our simulation, we explicitly verify that this correction disappears in the infinite system case. Third, we analyze the entanglement between two disjoint blocks, $B_1$ and $B_2$ separated by a segment containing a defect. Despite von-Neumann entropy's success in quantifying entanglement in CFTs, it does not quantify entanglement between subsystems when the total system~(in this case $B_1\cup B_2$) is in a mixed state, {\it i.e.}, when ${\rm Tr}\rho^2<1$. To that end, we consider the logarithmic entanglement negativity~(log-EN)~\cite{Vidal2002_EN,plenioLogarithmicNegativityFull2005} between the two blocks with the defect located symmetrically between them~[see Fig.~\ref{fig:geoms}(c)]: 
\begin{equation}
\label{eq:EN}
{\cal E}_{B_1,B_2} = \ln ||\rho^{T_{B_2}}|| = \ln {\rm Tr}|\rho^{T_{B_2}}|,
\end{equation}
where $||\rho^{T_{B_2}}||$ is the trace-norm given by the sum of the absolute values of the eigenvalues of $\rho^{T_{B_2}}$. We compute the log-EN using DMRG. The latter allows us to compute the log-EN directly for the spin-system. This is important since the log-EN for a spin system is not, in general, the same as that of a fermionic model obtained after Jordan-Wigner~(JW) transformation. This is because of the nonlocal nature of the JW transformation, where the fermionic degree of freedom in $B_2$ depends on the spin degrees of freedom in A~(see, for instance, Ref.~\cite{Igloi2010} for quantification of this effect for the EEs of disjoint intervals). We show that the log-EN also exhibits logarithmic scaling as a function of the separation of the blocks for reasons similar to the case of the EE. We determine the coefficient of this scaling as well as the subleading term from our simulations. The former defines an effective central charge which depends continuously on the defect strength. While there are no analytical results available for the log-EN in the presence of defects, we verify that we recover the expected results for the leading order term in the different extremal cases, while providing numerical results for the rest. Interestingly, the effective central charge obtained from the log-EN scaling is different from that obtained from the interface EE computations. Similar to the EEs, the duality defect, due to the existence of zero energy modes, manifests itself through an offset in the log-EN, which we determine numerically. Finally, we address the question of finite size effects in the log-EN due to the nonlocal zero energy mode present in the duality defect. We show that similar to the EE, the nonlocal zero energy mode leads to nontrivial finite size corrections for the log-EN. These lead to deviations from the logarithmic scaling that is expected in systems without zero energy modes.

The paper is organized as follows. Sec.~\ref{sec:defects} summarizes the lattice Hamiltonians for the different defects of the Ising CFT.  Sec.~\ref{sec:EE_Ising} describes the entanglement properties of the Ising model in the presence of defects. In Sec.~\ref{sec:EE_scaling}, we summarize the different scaling behaviors of the EE in the absence and presence of defects. This is followed by DMRG results of the symmetric and interface EE for open and infinite bcs~(Sec.~\ref{sec:symm_EE} and Sec.~\ref{sec:int_EE}). Sec.~\ref{sec:EN} discusses the log-EN for two disjoint blocks separated by a defect. We summarize the scaling behavior of the log-EN in Sec.~\ref{sec:EN_scaling}, followed by DMRG results in Sec.~\ref{sec:EN_res}. Sec.~\ref{sec:concl} provides a concluding summary and outlook. We note that the symmetric EE and the log-EN, unlike the interface EE, do not follow immediately from the DMRG simulations. Additional computation is necessary to extract the relevant quantities. The technical details of this computation are summarized in Appendix~\ref{sec:App_DMRG}.

\section{Defects in the Ising CFT}
\label{sec:defects}
In this section, we introduce the lattice Hamiltonians for the energy and duality defects in the Ising CFT for open and infinite bcs. The bcs are important since we are interested not only in the leading order scaling of the EE and EN, but also in the subleading terms. 

The energy defect constitutes altering a ferromagnetic coupling of two adjacent spins. The resultant Hamiltonian is:
\begin{equation}
\label{H_epsilon}
H_\epsilon = -\sum_{\substack{i=1\\ i\neq j}}^{L-1}\sigma_i^x\sigma_{i+1}^x - \sum_{i=1}^L\sigma_i^z-b_\epsilon\sigma_j^x\sigma_{j+1}^x,
\end{equation}
where $L$ is the length of the open chain and $b_\epsilon$ is the defect strength for a defect located between sites $j$ and $j+1$. Note that one can also view the open bc as another energy defect inserted between sites $L$ and 1 with zero defect strength. This model has been analyzed extensively in the past in the context of 2D classical Ising model with defect lines~(see, for example, Refs.~\cite{Kadanoff1971, Bariev1979, McCoy1980, Ko1985, Henkel1989, Baake1989, Grimm1990, Oshikawa1997}. The perturbing local defect term is marginal and gives rise to continuous scaling exponents~\cite{Cardy1987, Igloi1993}. The Casimir term of the ground state energy depends continuously on the defect strength~\cite{Henkel1989, Baake1989}:
\begin{equation}
\label{E0_def}
E_0(b_\epsilon) = -A_0L-A_1(b_\epsilon) - \frac{A_2(b_\epsilon)}{L} + \ldots,
\end{equation}
where $\ldots$ indicate subleading terms. Here, $A_0 = 4/\pi$ and 
\begin{equation}
\label{A2_def}
A_2(b_\epsilon) = \frac{\pi}{6}-2\pi[F(b_\epsilon) - F(0)]^2,\ F(b_\epsilon) = \frac{1}{4}-\frac{1}{\pi}\tan^{-1}b_\epsilon.
\end{equation}
The explicit form of $A_1$ is not relevant for our purposes. For $b_\epsilon = 1$, we have a homogeneous chain with open bc. For this case, we recover the well-known result of $A_2(1) = \pi/24$~[recall that with the conventions of Eq.~\eqref{H_epsilon}, the velocity of light is 2]. An antiferromagnetic defect~($b_\epsilon = -1$) yields the same result since the defect can be removed by applying a unitary transformation on one half of the chain. This is possible because we are dealing with an open chain\footnote{For a periodic ring with a single defect, $A_2(b_\epsilon) = \pi/6-2\pi F(b_\epsilon)^2$. In particular, for $b_\epsilon = 1$ (a periodic ring), we recover the standard result $A_2(1) = \pi/6$ and the expected central charge 1/2. On the other hand, for an antiperiodic chain, $A_2(-1) = -\pi/3$, yielding the effective central charge $-1$.~\cite{Grimm2001}}. For more on the algebraic properties for the Ising chain with multiple defects, we refer the reader to Refs.~\cite{Henkel1989, Baake1989, Grimm1992}. 
\begin{figure}
    \centering
    \includegraphics[width = \textwidth]{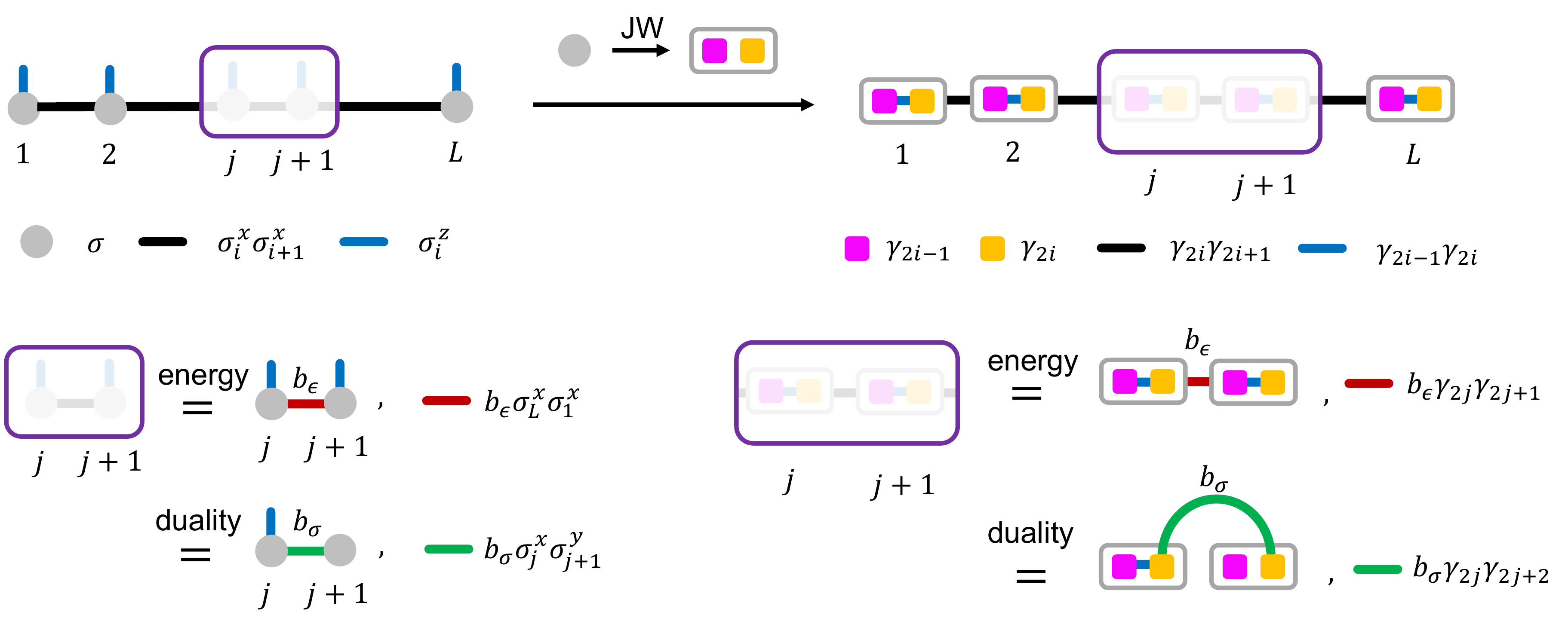}
    \caption{\label{def_schematic} Schematic of the defect Hamiltonians in the spin and fermionic pictures. The JW transformation maps each spin to a pair of Majorana fermions~[see Eq.~\eqref{eq:JW}]. The ferromagnetic couplings in the spin picture leads to couplings between Majorana fermions on neighboring sites~(black links), while the transverse field corresponds to couplings between Majoranas within each site~(blue links). The altered ferromagnetic coupling of the energy defect is indicated in maroon. The duality defect entails a $\sigma_{j}^x\sigma_{j+1}^y$ coupling indicated by a green bond, together with a missing transverse field on site $j+1$. In the fermionic picture, this leads to a Majorana zero mode~($\gamma_{2j+1}$). The other zero mode is delocalized throughout the system~[see Eq.~\eqref{eq:MZM_2}].}
\end{figure}

The duality defect consists of altering the ferromagnetic interaction of one bond to $b_\sigma\sigma_j^x\sigma_{j+1}^y$ where $b_\sigma$ is the interaction strength. Equally important, there is no transverse field at the $j+1^{\rm th}$ site. The Hamiltonian reads~\cite{Grimm2001, Oshikawa1997}
\begin{equation}
\label{H_sigma}
H_\sigma = -\sum_{\substack{i=1\\ i\neq j}}^{L-1}\sigma_i^x\sigma_{i+1}^x - \sum_{\substack{i=1\\i\neq j+1}}^L\sigma_i^z-b_\sigma\sigma_j^x\sigma_{j+1}^y. 
\end{equation}
The duality defect at $b_\sigma = 1$ is the topological defect in the Ising CFT~\footnote{We use the duality defect Hamiltonian of Refs.~\cite{Grimm2001, Roy2021a}, which is related by a local unitary rotation on the $j+1^{\rm th}$ spin to the one considered in Ref.~\cite{Oshikawa1997}.}. The nontrivial nature of the duality defect is manifest in the fermionic language after a JW transformation~\cite{Roy2021a}~(see Fig.~\ref{def_schematic}):
\begin{equation}\label{eq:JW}
\gamma_{2k-1} = \sigma_k^x\prod_{j=1}^{k-1}\sigma_j^z, \gamma_{2k} = \sigma_k^y\prod_{j=1}^{k-1}\sigma_j^z,
\end{equation}
where $\gamma_{j}$-s are real, Majorana fermion operators obeying $\{\gamma_j, \gamma_k\} = 2\delta_{j,k}$. In the fermionic language, the duality defect Hamiltonian is
\begin{equation}
H_\sigma^f =  \frac{i}{2}\sum_{\substack{i=1\\i\neq j}}^{L-1}\gamma_{2i}\gamma_{2i+1} + \frac{i}{2}\sum_{\substack{i=1\\i\neq j+1}}^{L}\gamma_{2j-1}\gamma_{2j}+\frac{ib_\sigma}{2}\gamma_{2j}\gamma_{2j+2}.
\end{equation}
The fermionic formulation also clarifies that the duality defect part of the Hamiltonian~(the last term in the previous equation) is also a marginal perturbation and thus, gives rise to continuously varying scaling exponents.  Note that the operator $\gamma_{2j+1}$ does not occur in $H_\sigma^f$. It commutes with the Hamiltonian and anticommutes with the conserved $\mathbb{Z}_2$ charge $Q$: 
\begin{equation}
\label{eq:MZM_1}
[\gamma_{2j+1}, H_\sigma^f] = 0, \ \{\gamma_{2j+1},Q\} =0,\ Q = \prod_{i=1}^L(-i\gamma_{2j-1}\gamma_{2j}).
\end{equation}
Thus, it is a zero-mode of the model which is perfectly localized in space. It has a partner zero-mode which is completely delocalized: 
\begin{equation}
\label{eq:MZM_2}
\Lambda(b_\sigma) = b_\sigma\sum_{k=1}^{j}\gamma_{2k-1} + \sum_{k = j+1}^{L}\gamma_{2k}.
\end{equation}
Note that the zero-modes exist for all values of $b_\sigma$ and are not special features of the topological point. The fermionic Hamiltonian also reaffirms a CFT result~\cite{Grimm2001}:~$H_\sigma^f$ describes a chain of $2L-1$ Majorana fermions or equivalently, $L-1/2$ spins. This is important for quantifying finite-size effects~\cite{Roy2021a}. 

\section{Entanglement Entropies in the Ising CFT with defects}
\label{sec:EE_Ising}
In this section, we briefly summarize the known properties of the EE in the Ising CFT for the two geometries considered in Fig.~\ref{fig:geoms}(a,b). First, we describe the logarithmic scaling of the EE in the absence of defects. Subsequently, we discuss the modifications of the aforementioned behavior of the EE due to the existence of defects. This is followed by a presentation of numerical results obtained using DMRG. All lengths are measured in units of the lattice spacing and are dimensionless.

\subsection{Scaling of Entanglement Entropies}
\label{sec:EE_scaling}
In the absence of defects, the behavior of the EE can be inferred from standard computations based on the replica trick~\cite{Holzhey1994, Calabrese2004, Calabrese2009}. For an infinite system, the EE for a subsystem of size~$l$ is given by
\begin{equation}
\label{eq:S_block}
{\mathcal S}(l) = \frac{c+c}{6}\ln l + S_0=\frac{c}{3}\ln l + S_0,
\end{equation}
where~$c=1/2$ the central charge of the Ising model and $S_0$ is a nonuniversal constant. Here, in the first equality, we have explicitly shown that each edge of the subsystem~(the entanglement cuts) contribute a factor of $c/6$ to the leading logarithmic dependence. For a finite system of length~$L$, in the absence of zero-energy modes, the dependence is similar, with~$l$ replaced by the chord length $L\sin(\pi l/L)/\pi$~\cite{Calabrese2004, Calabrese2009}. The presence of the defect changes this dependence. This is explained below. 

Consider first the case when the defect is {\it not} at the boundary between the subsystem and the rest~(of an infinite system). For simplicity, we consider the case when the subsystem is located symmetrically around the defect, {\it i.e.}, the case of the symmetric EE~(${\cal S}_\mathrm{S}$). In this case, the leading order logarithmic dependence is not altered. This is because the leading logarithmic contribution, which arises from the correlations around the entanglement cut, is {\it totally oblivious} of the defect. The defect strength could even be chosen such that the chain is cut into half without changing the leading logarithmic scaling. However, the subleading~${\cal O}(1)$ contribution contains information about the defect. This becomes apparent in the after folding the system at the defect, thereby transforming the defect problem to a boundary CFT one. Then, the subleading~${\cal O}(1)$ term can be equated to a boundary entropy with double the bulk degrees of freedom~\cite{Oshikawa1997, Saleur1998, Saleur2000}. The resulting expression is
\begin{equation}
\label{eq:S_block_1}
{\mathcal S}_\mathrm{S}(l) =\frac{c}{3}\ln l + \tilde{S}_0,
\end{equation}
where we have indicated the modified subleading expression by $\tilde{S}_0$. For a finite-size system, as for without any defects, the scaling behavior is obtained by replacing the subsystem-length by the chord-length. 

Now, consider the case when the defect lies exactly at one of the edges of the subsystem, {\it i.e.}, the case of the interface EE. The other end of the subsystem is somewhere in the bulk of the total system. It turns out that for the Ising CFT with defects, the interface EE, also exhibits a logarithmic scaling. But, both the leading order logarithmic scaling and the subleading~${\cal O}(1)$ terms are modified:  
\begin{equation}
\label{eq:S_int_1}
{\mathcal S}_\mathrm{I}(l) = \frac{c_{\rm eff} + c}{6}\ln l + S'_0,
\end{equation}
where~$c_{\rm eff}$ is an effective central charge that depends continuously on the defect strength with~$c_{\rm eff}\leq c$. Note that this equation is valid for an infinite total system size. The coefficients of the logarithm, $c_{\rm eff}$ and $c$, arise from the entanglement cuts at the defect and the other end of the subsystem respectively. This logarithmic scaling and the continuous dependence of~$c_{\rm eff}$ on the defect strength is a non-generic scenario that arises due to the fact that the defect terms in the Hamiltonian are marginal perturbations, which gives rise to continuous scaling exponents. It is also worth noting that the effective central charge is merely the coefficient of the logarithmic scaling and bears little algebraic significance. Importantly,~$c_{\rm eff}$ is different from the central charge obtained from the ground state energy dependence~(see Sec.~\ref{sec:defects}). Note also that the subleading term,~$S'_0$, is also different compared to the case without defects. For a finite total system of length $L$, often one can consider the interface EE between the left and right halves of the system. Then, the interface EE is
\begin{equation}
\label{eq:S_int_2}
{\mathcal S}_\mathrm{I}(L) = \frac{c_{\rm eff}}{6}\ln L + S''_0.
\end{equation}
Note that there is no factor of $c$ in the coefficient arising from the second entanglement cut, which now coincides with the physical boundary of the system. Also, the subleading term is different compared to Eq.~\eqref{eq:S_int_1} since now it depends explicitly on the bcs of the system.~\footnote{\label{foot_int_EE}In the absence of a defect, the EE between the left~(of length $l$) and the right~(of length $L-l$) parts of the system is given by~\cite{Calabrese2004, Calabrese2009}: ${\mathcal S}(l) = \frac{c}{6}\ln\big(\frac{2L}{\pi}\sin\frac{\pi l}{L}\big) + \ldots$, where the dots indicate the subleading contributions. In the absence of the defect, Eq.~\eqref{eq:S_int_2} is a special case: $l = L/2$. However, in the presence of a defect, only the interface EE between left and right halves is known for finite systems.}

In this way, both the symmetric and interface EEs serve as viable diagnostics for detection of defects. The case of the interface EE is particularly interesting since there is no simple folding maneuver that converts the defect problem to a boundary CFT one. Clearly, in the absence of a defect, the $c_{\rm eff} = c$, which can be viewed as a manifestation of perfect transmission of modes carrying information. In the presence of a defect, reflection of these information-carrying modes causes $c_{\rm eff}$ to be less than $c$. However, there exist nontrivial defects, which still leave $c_{\rm eff}=c$. Topological defects are precisely such perfectly-transmissive defects, which leave their fingerprints only in the subleading term, $S'_0$ or $S''_0$, in the interface EE (see Sec.~\ref{sec:DMRG_res}). 

The preceding discussion about the logarithmic dependence of the symmetric and interface EEs is applicable only in the absence of zero energy modes. Defects in the Ising CFT can give rise to both local and nonlocal zero energy modes: {\it e.g.}, the critical Ising chain with antiperiodic bc~\cite{Cardy1984} and the Ising CFT with duality defects~(see Sec.~\ref{sec:defects}). In the computation of ground-state properties, one can consider the case when the zero energy modes are filled or empty. The system could also be considered in a statistical mixture of the filled and empty states. The purity of the system has nontrivial consequences on the entanglement properties of its subsystems~\cite{Herzog2013, Klich2017}. For Hamiltonians with zero energy modes, the logarithmic dependence of the EE is valid, in general, only in the limit when subsystem size is infinitesimal compared to the system size. When the subsystem occupies a finite fraction of the system, the nonlocal zero energy modes give rise to nontrivial finite size corrections~\cite{Klich2017, Roy2021a}. The latter are particularly important for the interface EEs between the left and right halves of the system, where the subsystem is necessarily a finite fraction of the total system. Next, we consider the case when the total state of the system is pure~(see Ref.~\cite{Roy2021a} for results when the total system is mixed) and present numerical results for the symmetric and interface EEs. 

\subsection{DMRG Results}
\label{sec:DMRG_res}
Here, we consider the various defect Hamiltonians of Sec.~\ref{sec:defects} and compute the EEs for different subsystems using DMRG. The DMRG simulations were performed using the TeNPy package~\cite{Hauschild2018}. We consider two cases: i) where the size of the total system is infinite and ii) when the size of the total system is finite with open bcs at the ends. We consider these two different cases to explicitly demonstrate the influence of the non-local zero mode for the topological defect of the Ising CFT. For technical details of the computation, see Appendix~\ref{sec:App_DMRG}. 
\subsubsection{Symmetric Entanglement Entropy}
\label{sec:symm_EE}
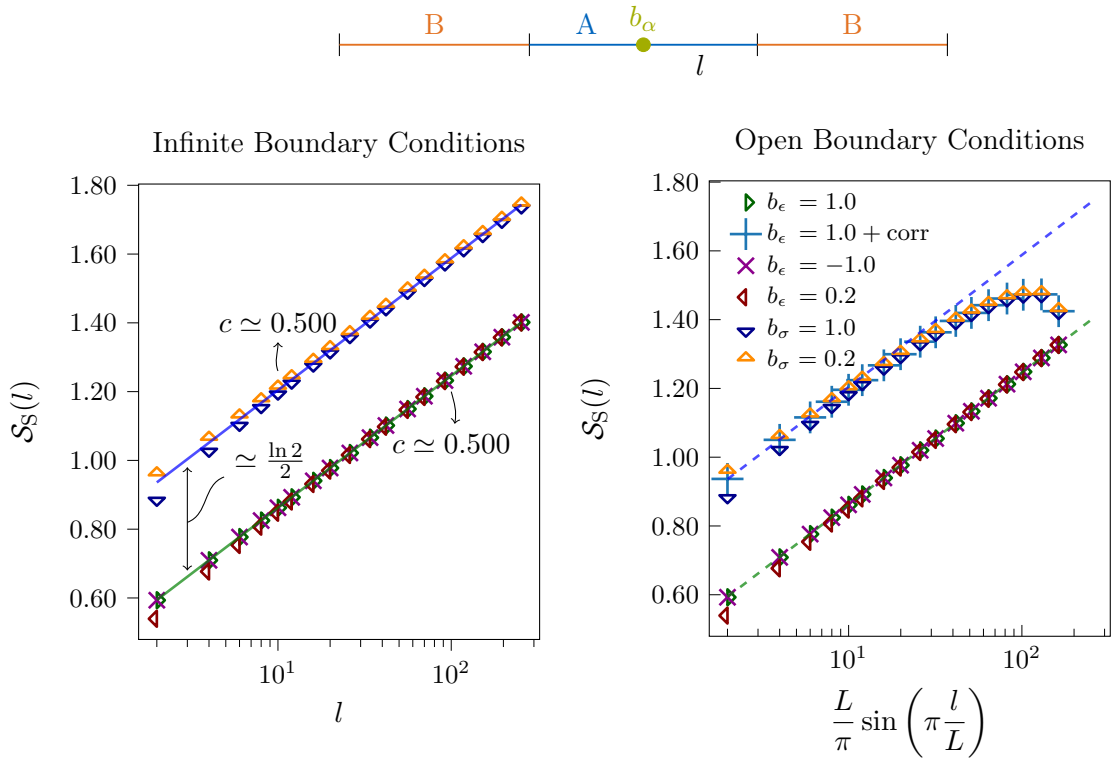
\begin{figure}
    \begin{minipage}{\linewidth}
    \centering
    \hspace{1.8cm}
    \begin{tikzpicture}
    \definecolor{color0}{rgb}{0,0.396078431372549,0.741176470588235}
    \definecolor{color1}{rgb}{0.890196078431372,0.447058823529412,0.133333333333333}
    \definecolor{color2}{rgb}{0.635294117647059,0.67843137254902,0}
    \draw
        (0,0.15) -- (0, -0.15);
    \draw
        [line width = 0.25mm, color1]
        (0,0)--(2.5,0)
        node[midway, above]{B};
    \draw
        (2.5,0.15) -- (2.5, -0.15);
    \draw
        [line width = 0.25mm, color0]
        (2.5,0) -- (5.5,0)
        node[pos=0.25,above, color0] {A}
        node[pos = 0.75,below,black] {$l$};
    \draw
        (5.5,0.15) -- (5.5, -0.15);
    \draw
        [line width = 0.25mm, color1] (5.5,0) -- (8,0)
        node[midway, above]{B};
    \draw
        (8,0.15) -- (8, -0.15);
    \node[circle,
    		draw = color2,
    		fill = color2,
    		scale = 0.5] (d) at (4,0) {};
	\node[text=color2, above] (d) at (4,0.05) {$b_\alpha$};

\end{tikzpicture}
    \end{minipage}
    \begin{minipage}{.45\linewidth}
\begin{tikzpicture}[
baseline=(current bounding box.center)
]

\definecolor{color0}{rgb}{0.545098039215686,0,0.545098039215686}
\definecolor{color1}{rgb}{1,0.549019607843137,0}

\begin{axis}[
height=3.0in,
log basis x={10},
minor xtick={0.2,0.3,0.4,0.5,0.6,0.7,0.8,0.9,2,3,4,5,6,7,8,9,20,30,40,50,60,70,80,90,200,300,400,500,600,700,800,900,2000,3000,4000,5000,6000,7000,8000,9000,20000,30000,40000,50000,60000,70000,80000,90000},
minor ytick={},
tick align=outside,
tick pos=left,
title={Infinite Boundary Conditions},
width=2.7in,
x grid style={white!69.0196078431373!black},
xlabel={\(\displaystyle l\)},
xmin=1.56978367970064, xmax=323.611467343625,
xmode=log,
xtick style={color=black},
xtick={0.1,1,10,100,1000,10000},
y grid style={white!69.0196078431373!black},
ylabel={\(\displaystyle \mathcal{S}_\mathrm{S}(l)\)},
ymin=0.479304033055729, ymax=1.80362627464997,
ytick style={color=black},
ytick={0.4,0.6,0.8,1,1.2,1.4,1.6,1.8,2}
]
\addplot [
  draw=green!39.2156862745098!black,
  fill=green!39.2156862745098!black,
  mark=Custom_9,
  only marks
]
table{%
x  y
2 0.593397192709205
4 0.709459700280549
6 0.777114100020172
8 0.825087137614272
10 0.862289635913136
12 0.89268299453142
16 0.940636380020272
20 0.977829916178975
26 1.02155943174177
34 1.0662713866918
42 1.10149019945531
56 1.14943774415013
70 1.18662858394311
92 1.23217765020673
118 1.27366041466719
152 1.3158597336548
196 1.35823188775747
254 1.40143470483231
};
\addplot [draw=color0, fill=color0, mark=x, only marks]
table{%
x  y
2 0.593397192141813
4 0.709459699646506
6 0.777114099350178
8 0.825087136894234
10 0.862289635184872
12 0.892682993801524
16 0.940636379246894
20 0.977829915165005
26 1.02155943032654
34 1.06627138498151
42 1.10149019730869
56 1.14943774092746
70 1.18662857885385
92 1.23217764222119
118 1.27366040276396
152 1.31585971765781
196 1.35823186769204
254 1.4014346750058
};
\addplot [
  draw=red!54.5098039215686!black,
  fill=red!54.5098039215686!black,
  mark=Custom_8,
  only marks
]
table{%
x  y
2 0.539500498582741
4 0.676943556036003
6 0.754104252213603
8 0.807328763282782
10 0.84784475145752
12 0.880514865952652
16 0.9313899672537
20 0.970376598491511
26 1.01578769314707
34 1.06183618388863
42 1.09789019029452
56 1.1467318960585
70 1.18446291300675
92 1.23053198805733
118 1.27238238743971
152 1.31487645047289
196 1.35748392210057
254 1.40088166858339
};
\addplot [
  draw=blue!54.5098039215686!black,
  fill=blue!54.5098039215686!black,
  mark=Custom_11,
  only marks
]
table{%
x  y
2 0.891155283019918
4 1.03365482941785
6 1.10904520696002
8 1.16072007114936
10 1.20008178276475
12 1.23188106277888
16 1.2815356130507
20 1.31969978276945
26 1.36426055839182
34 1.40953687667783
42 1.44502604837608
56 1.49312008652445
70 1.53025639397627
92 1.57553519899858
118 1.61656116282381
152 1.65806036688723
196 1.69944628227225
254 1.7412860797147
};
\addplot [draw=color1, fill=color1, mark=Custom_10, only marks]
table{%
x  y
2 0.956803007022249
4 1.06016190940408
6 1.12514562624392
8 1.1721862277088
10 1.20895777372062
12 1.23911099624048
16 1.28679935943244
20 1.32383240819823
26 1.36738083918761
34 1.41188356354399
42 1.44690195688055
56 1.49450099040924
70 1.53134132529832
92 1.57633529562493
118 1.61715778455135
152 1.65849039884778
196 1.69973930980047
254 1.74146229550368
};
\addplot [line width=1pt, green!50.1960784313725!black, opacity=0.7]
table {%
2 0.594050458267005
4 0.709578297275395
6 0.777157750884654
8 0.825106136283785
10 0.86229779340221
12 0.892685589893044
16 0.940633975292176
20 0.9778256324106
26 1.02155426228466
34 1.06626620734375
42 1.10148539835954
56 1.14943378375867
70 1.18662544087709
92 1.23217563467923
118 1.27365949741751
152 1.31586002217621
196 1.35823359222516
254 1.40143809764082
};
\addplot [line width=1pt, blue, opacity=0.7]
table {%
2 0.935992853559366
4 1.05152774914449
6 1.11911133058653
8 1.16706264472962
10 1.20425657355835
12 1.23464622617165
16 1.28259754031475
20 1.31979146914348
26 1.36352277001385
34 1.40823744613131
42 1.44345878837684
56 1.49141010251994
70 1.52860403134867
92 1.57415700741039
118 1.61564340403196
152 1.65784650644851
196 1.70022266472513
254 1.74342980912296
};
\coordinate (topLine) at (axis cs: 10, 1.26);
\coordinate (botLine) at (axis cs: 100, 1.2);
\node (c1) at (axis cs:10,1.4) {$c \simeq 0.500$};
\node (c2) at (axis cs:100,1.05) {$c \simeq 0.500$};
\draw[<-] (c1) to[out = -90, in = 110] (topLine);
\draw[<-] (c2)  to[out = 80, in = -75] (botLine);
\coordinate (offsetTop) at (axis cs: 3, 0.98);
\coordinate (offsetBot) at (axis cs: 3, 0.68);
\coordinate (offsetMid) at (axis cs: 3, 0.82);
\draw[<->] (offsetTop) -- (offsetBot);
\node (offsetLabel) at (axis cs:9, 1) {$\simeq \frac{\ln{2}}{2}$};
\draw[-] (offsetLabel) to[out = -160, in = 20] (offsetMid);
\end{axis}

\end{tikzpicture}
    \end{minipage}
    \begin{minipage}{.45\linewidth}
        \vspace{12 pt}
        \hspace{14 pt}
\begin{tikzpicture}[
baseline=(current bounding box.center)
]

\definecolor{color0}{rgb}{0.12156862745098,0.466666666666667,0.705882352941177}
\definecolor{color1}{rgb}{0.545098039215686,0,0.545098039215686}
\definecolor{color2}{rgb}{1,0.549019607843137,0}

\begin{axis}[
height=3.0in,
legend cell align={left},
legend style={
  fill opacity=0.8,
  draw opacity=1,
  text opacity=1,
  at={(0.03,1.0)},
  anchor=north west,
  draw=none
},
log basis x={10},
minor xtick={0.2,0.3,0.4,0.5,0.6,0.7,0.8,0.9,2,3,4,5,6,7,8,9,20,30,40,50,60,70,80,90,200,300,400,500,600,700,800,900,2000,3000,4000,5000,6000,7000,8000,9000,20000,30000,40000,50000,60000,70000,80000,90000},
minor ytick={},
tick pos=left,
title={Open Boundary Conditions},
width=2.7in,
x grid style={white!69.0196078431373!black},
xlabel={\(\displaystyle \dfrac{L}{\pi}\sin\left(\pi\dfrac{l}{L}\right)\)},
xmin=1.56974230888051, xmax=323.611873473488,
xmode=log,
xtick style={color=black},
xtick={0.1,1,10,100,1000,10000},
y grid style={white!69.0196078431373!black},
ylabel={\(\displaystyle \mathcal{S}_\mathrm{S}(l)\)},
ymin=0.479223722978806, ymax=1.80363009893935,
ytick style={color=black},
ytick={0.4,0.6,0.8,1,1.2,1.4,1.6,1.8,2}
]
\addplot [
  draw=green!39.2156862745098!black,
  fill=green!39.2156862745098!black,
  mark=Custom_9,
  only marks
]
table{%
x  y
1.9999498009741 0.592657690673474
3.99959841686457 0.708693836432142
5.99864470794513 0.776324336583088
7.99678762519743 0.824266955262577
9.99372625564787 0.861431224394082
11.989159867684 0.891778199867709
15.9743102885699 0.939613932339623
19.9498383822559 0.976656384423262
25.8898521722725 1.02009632617031
31.7947792366146 1.05433861528168
41.5366446596453 1.09888391448141
51.1221755311639 1.13348985616183
64.2107194517257 1.17147941893922
82.0639977538064 1.21236273831194
101.83626370635 1.24833244433451
129.70125031778 1.28862393609839
162.962390033807 1.32656106110563
};
\addlegendentry{$b_{\epsilon \:} = 1.0$}
\addplot [draw=color0, fill=color0, mark=+, only marks]
table{%
x  y
1.9999498009741 0.936824849896686
3.99959841686457 1.05046085675558
5.99864470794513 1.11569752746855
7.99678762519743 1.16125264355705
9.99372625564787 1.19603575441057
11.989159867684 1.22400793346647
15.9743102885699 1.26711322928153
19.9498383822559 1.29945090476674
25.8898521722725 1.33588210816487
31.7947792366146 1.36317420637978
41.5366446596453 1.39626731088935
51.1221755311639 1.4195874129425
64.2107194517257 1.44206085152009
82.0639977538064 1.46136551298081
101.83626370635 1.47237912198091
129.70125031778 1.47344978742466
162.962390033807 1.42457229000032
};
\addlegendentry{$b_{\epsilon \:} = 1.0 + \mathrm{corr} $}
\addplot [draw=color1, fill=color1, mark=x, only marks]
table{%
x  y
1.9999498009741 0.592657690673258
3.99959841686457 0.708693836432094
5.99864470794513 0.776324336582758
7.99678762519743 0.824266955262366
9.99372625564787 0.861431224393736
11.989159867684 0.891778199867634
15.9743102885699 0.939613932339557
19.9498383822559 0.976656384424642
25.8898521722725 1.02009632616949
31.7947792366146 1.05433861528105
41.5366446596453 1.09888391447906
51.1221755311639 1.13348985616289
64.2107194517257 1.17147941892862
82.0639977538064 1.21236273830973
101.83626370635 1.24833244433874
129.70125031778 1.28862393610518
162.962390033807 1.32656106069908
};
\addlegendentry{$b_{\epsilon \:} = -1.0$}
\addplot [
  draw=red!54.5098039215686!black,
  fill=red!54.5098039215686!black,
  mark=Custom_8,
  only marks
]
table{%
x  y
1.9999498009741 0.539424012795195
3.99959841686457 0.676846644582668
5.99864470794513 0.753980744134008
7.99678762519743 0.807170847788408
9.99372625564787 0.847644291568263
11.989159867684 0.88026361484242
15.9743102885699 0.931012220839452
19.9498383822559 0.969839007577714
25.8898521722725 1.0149476121469
31.7947792366146 1.05024106045174
41.5366446596453 1.09587383744946
51.1221755311639 1.13114721353535
64.2107194517257 1.16972620466153
82.0639977538064 1.21110922922682
101.83626370635 1.24742748104281
129.70125031778 1.28803477927138
162.962390033807 1.32628962213131
};
\addlegendentry{$b_{\epsilon \:} = 0.2$}
\addplot [
  draw=blue!54.5098039215686!black,
  fill=blue!54.5098039215686!black,
  mark=Custom_11,
  only marks
]
table{%
x  y
1.9999498009741 0.889239993298084
3.99959841686457 1.02944533903108
5.99864470794513 1.10249280722698
7.99678762519743 1.1518097294684
9.99372625564787 1.18880642219621
11.989159867684 1.21823619652473
15.9743102885699 1.26314209765061
19.9498383822559 1.29654731583318
25.8898521722725 1.33395301029306
31.7947792366146 1.36184659932733
41.5366446596453 1.3955501955616
51.1221755311639 1.41923703759433
64.2107194517257 1.44202620879514
82.0639977538064 1.46158713255503
101.83626370635 1.47276550711582
129.70125031778 1.47395820982685
162.962390033807 1.42508753128605
};
\addlegendentry{$b_{\sigma} = 1.0$}
\addplot [draw=color2, fill=color2, mark=Custom_10, only marks]
table{%
x  y
1.9999498009741 0.954885937191549
3.99959841686457 1.05594983512321
5.99864470794513 1.11858985665032
7.99678762519743 1.16327174556263
9.99372625564787 1.19767751776399
11.989159867684 1.22546049547109
15.9743102885699 1.26839877665031
19.9498383822559 1.30067149992189
25.8898521722725 1.33706289263815
31.7947792366146 1.36433681196574
41.5366446596453 1.39741105970138
51.1221755311639 1.42071429134959
64.2107194517257 1.44316236195868
82.0639977538064 1.46242681821013
101.83626370635 1.473388234295
129.70125031778 1.47436735126641
162.962390033807 1.42524236738732
};
\addlegendentry{$b_{\sigma} = 0.2$}
\addplot [line width=1pt, green!50.1960784313725!black, opacity=0.7, dashed, forget plot]
table {%
2 0.594050458267005
4 0.709578297275395
6 0.777157750884654
8 0.825106136283785
10 0.86229779340221
12 0.892685589893044
16 0.940633975292176
20 0.9778256324106
26 1.02155426228466
34 1.06626620734375
42 1.10148539835954
56 1.14943378375867
70 1.18662544087709
92 1.23217563467923
118 1.27365949741751
152 1.31586002217621
196 1.35823359222516
254 1.40143809764082
};
\addplot [line width=1pt, blue, opacity=0.7, dashed, forget plot]
table {%
2 0.935992853559366
4 1.05152774914449
6 1.11911133058653
8 1.16706264472962
10 1.20425657355835
12 1.23464622617165
16 1.28259754031475
20 1.31979146914348
26 1.36352277001385
34 1.40823744613131
42 1.44345878837684
56 1.49141010251994
70 1.52860403134867
92 1.57415700741039
118 1.61564340403196
152 1.65784650644851
196 1.70022266472513
254 1.74342980912296
};
\end{axis}

\end{tikzpicture}
    \end{minipage}
    \caption{\label{fig:symm_EE}Scaling of the symmetric EE, $\mathcal{S}_\mathrm{S}(l)$, for a segment A of length~$l$ located symmetrically around a defect. We show the results for infinite (left) and open (right) bcs. For finite bc, the total system length is denoted by $L$. (Left) The symmetric EE scales logarithmically with $l$ for both energy and duality defects for all defect strengths~[see Eq.~\eqref{eq:S_block_1}]. The coefficient of the logarithm for every case gives the expected central charge $c = 1/2$ (the errors in the fits occur in the fourth decimal places). Recall that this coefficient is sensitive to correlations around the entanglement cut, which in this case is located far from the defect~(see discussions in Sec.~\ref{sec:EE_scaling}). The subleading is sensitive to the defect type. For the duality defect of a given strength, this subleading term is $(\ln2)/2$ higher than the energy defect of the corresponding strength. This can be viewed as a consequence of the zero-energy mode localized at the defect~[see Eq.~\eqref{eq:MZM_1}]. We note that the symmetric EE does not seem to depend, at least up to ${\cal O}(1)$ corrections, to the actual defect strength value for a given defect class. (Right) For finite systems, the results are similar to the infinite case for $l\ll L$ with the subsystem size replaced by the chord-length. As $l$ is increased to values comparable to $L$, deviations occur from usual logarithmic scaling for the duality defect cases due to the existence of the delocalized zero-energy mode~[see Eq.~\eqref{eq:MZM_2}]. The predicted values of the duality defect cases are indicated by the blue crosses, which are obtained by adding $\Delta S(1-l/L)/2$ to the symmetric EE without any defect. }
\end{figure}

Consider the EE of a subsystem located symmetrically around the energy defect for an infinite system~(left panel of Fig.~\ref{fig:symm_EE}). The DMRG simulations were performed using the single-site DMRG algorithm for different defect strengths~$b_\epsilon$. The maximum bond dimension~($\chi$) used was $100$. While the ground state can be obtained, in general, for much larger values of $\chi$, the computation of the symmetric EE~(and the log-EN performed later) scales as $\chi^4$ and thus, prevents us from using larger values of $\chi$. However, as shown below, this low bond-dimension is already sufficient to capture the universal properties of the model. 
The state was obtained by converging the relative ground state energy to $10^{-11}$ for and the average von-Neumann entropy up to $10^{-8}$ for the chosen $\chi = 100$. Recall that $b_\epsilon = +1$ corresponds a homogeneous infinite chain without any defect. For both the energy defects of different strengths, as expected from Eq.~\eqref{eq:S_block_1}, the symmetric EE scales with $\ln l$, $l$ being the system size with a coefficient~$c/3$, where $c = 1/2$. The subleading term~$\tilde{S}_0$, consisting of universal and non-universal parts, is also independent of the defect strength~$b_\epsilon$. In the folded picture, the universal part corresponds to the same boundary entropy due to the same $g$-function~($=1$)~\cite{Oshikawa1997, Bachas2013}. The specific case of $b_\epsilon = -1$ can be verified explicitly since the defect can be removed by a unitary transformation on one half of the infinite chain. Similar computations were done for the finite system~(right panel of Fig.~\ref{fig:symm_EE}). The results are identical after substituting the $l\rightarrow L\sin(\pi l/L)/(\pi)$. 

The duality defect on the other hand shows dramatically different behavior. As expected, the leading logarithmic scaling is identical to the case of the energy defect for the same reasoning. However, the subleading term, $\tilde{S}_0$, is different. For {\it all} subsystem sizes in an infinite system and for subsystem sizes $l\ll L$ in the finite system, 
\begin{equation}
\tilde{S}_0 = S_0 + \frac{1}{2}\ln 2
\end{equation}
This result is true for all strengths,~($b_\sigma$), of the duality defect and not just the topological point:~$b_\sigma=1$. Note that only the difference between $\tilde{S}_0$ and $S_0$ is universal. Physically, this extra contribution arises due the presence of the {\it localized} unpaired Majorana zero mode localized within the subsystem~[see Eq.~\eqref{eq:MZM_1}]. In the folded picture, this defect corresponds to the `continuous Neumann boundary fixed point'~\cite{Oshikawa1997}, with a $g$-function $=\sqrt{2}$. The difference in the symmetric EEs between the duality and the energy defects is given precisely by difference of the logarithm of the two $g$-functions. As shown in Fig.~\ref{fig:symm_EE}~(right panel), for a finite system size $L$, when subsystem occupies an appreciable fraction of the total system, the symmetric EE no longer exhibits the usual logarithmic scaling. This is because of the delocalized Majorana zero mode~[see Eq.~\eqref{eq:MZM_2}] present in the system. The complete functional dependence of the subleading term is~\cite{Roy2021a}:
\begin{equation}
\label{eq:symm_EE_offset}
\tilde{S}_0 = S_0 + \frac{1}{2}\Delta S\Big(1-\frac{l}{L}\Big),
\end{equation}
where~\cite{Klich2017}
\begin{equation}
\label{eq:dS_K}
\Delta S\Big(\frac{l}{L}\Big) = \frac{\pi l}{L}\int_0^\infty dh \tanh\Big(\frac{\pi l h}{L}\Big)[\coth(\pi h) - 1],
\end{equation}
Note that $l/L\ll 1$, $\Delta S\sim \pi^2l^2/12L^2$, while $\Delta S(1) = \ln2$. This nontrivial expression for the EE was first computed for an antiperiodic Ising chain in Ref.~\cite{Klich2017}. In contrast to the latter problem, which had {\it two} nonlocal zero modes, the Ising model with duality defects contains one local and {\it one} nonlocal zero mode. This accounts for the factor of 1/2 difference between our result and those obtained in Ref.~\cite{Klich2017}. 

\subsubsection{Interface Entanglement Entropy}
\label{sec:int_EE}
Next, we investigate the scaling of the interface EE. We compute the EE for different bipartionings of an infinite system and a finite system with open bc. The defect is located at the center of the system and thus, one specific choice of partitioning yields the interface EE for the model. For the scaling analysis, we consider only the interface EE between the left and right halves of the infinite/finite system. In this case, for a finite system, the scaling behavior of the interface EE is given Eq.~\eqref{eq:S_int_2}. For an infinite system, the scaling is more nontrivial. In principle, for the infinite system in the presence of defects, the correlation length is still infinite~(recall that the defect perturbations are marginal perturbations of the Ising CFT). However, in an actual DMRG simulation of an infinite system, the correlation length is rendered finite due to `finite entanglement truncation'~\cite{Tagliacozzo2008, Pollmann2009}. The entanglement truncation is a direct consequence of the finite number of Schmidt states kept during the numerical simulation.  The resulting finite correlation length provides a natural length-scale for the scaling of the interface EE. In the absence of defects, this scaling is analogous to the finite-size scaling, with the system size being replaced by the computed finite correlation length in Eq.~\eqref{eq:S_int_2}. In this work, we show that the scaling obtained due to finite entanglement truncation works~(surprisingly well) for the case with defects, with the correlation length computed for the system without defects keeping the entanglement truncation the same. We also note that this issue arises only for the scaling of the interface EE between left and right halves of an infinite system. In the other cases, the block size provides the relevant length-scale for scaling~(see Sec.~\ref{sec:EE_scaling}).

\begin{figure}
    \centering
    \hspace{1.6cm}
\begin{tikzpicture}
    \definecolor{color0}{rgb}{0,0.396078431372549,0.741176470588235}
    \definecolor{color1}{rgb}{0.890196078431372,0.447058823529412,0.133333333333333}
    \definecolor{color2}{rgb}{0.635294117647059,0.67843137254902,0}

    \draw[line width=0.25mm, color0 ]  (0,0) -- (2.5,0) node[midway,above] {A};
    \draw[line width=0.25mm, color1 ]  (2.5,0) -- (8,0) node[midway,above] {B};
\node[circle,
    draw = color2,
    fill = color2,
    scale = 0.5] (d) at (4,0) {};
\node[text=color2, above] (d) at (4,0.05) {$b_\alpha$};
    \draw (2.5,0.15) -- (2.5, -0.15)node[below]{$l$};
    \draw (0,0.15) -- (0, -0.15);
    \draw (8,0.15) -- (8, -0.15);
\end{tikzpicture}
    \\
    \hspace{-4em}
    \begin{minipage}{.45\linewidth}
        \begin{tikzpicture}[
  baseline=(current bounding box.center)
  ]
  
  \definecolor{color0}{rgb}{0.545098039215686,0,0.545098039215686}
  \definecolor{color1}{rgb}{1,0.549019607843137,0}
  
  \begin{axis}[
  height=3.0in,
  minor xtick={},
  minor ytick={},
  tick pos=left,
  title={Infinite Boundary Conditions},
  width=2.7in,
  x grid style={white!69.0196078431373!black},
  xlabel={\(\displaystyle l\)},
  xmin=-24.5, xmax=536.5,
  xtick style={color=black},
  xtick={-200,0,200,400,600},
  y grid style={white!69.0196078431373!black},
  ylabel={\(\displaystyle \mathcal{S}(l)\)},
  ymin=0.377569415512266, ymax=1.16286114060767,
  ytick style={color=black},
  ytick={0.3,0.4,0.5,0.6,0.7,0.8,0.9,1,1.1,1.2}
  ]
  \addplot [
    draw=green!39.2156862745098!black,
    fill=green!39.2156862745098!black,
    mark=Custom_9,
    only marks
  ]
  table{%
  x  y
  1 1.05301454833048
  13 1.05301480177125
  26 1.05301506820808
  38 1.05301530365643
  51 1.05301554589344
  64 1.05301577444471
  77 1.05301598872861
  89 1.05301617536539
  102 1.05301636430377
  115 1.05301653887016
  128 1.05301669885035
  141 1.05301684462537
  153 1.05301696635292
  166 1.05301708492169
  179 1.05301718865169
  192 1.05301727788727
  204 1.05301734660261
  217 1.05301740711271
  230 1.05301745265002
  243 1.05301748310616
  255 1.05301749722433
  268 1.05301749815959
  281 1.0530174846886
  294 1.05301745642743
  307 1.05301741393901
  319 1.05301736042884
  332 1.05301728712107
  345 1.05301719889344
  358 1.05301709566958
  370 1.05301698704152
  383 1.05301685557914
  396 1.05301670942984
  409 1.05301654887541
  422 1.05301637408926
  434 1.05301620004502
  447 1.05301599828792
  460 1.05301578284796
  473 1.05301555412387
  485 1.05301533054571
  498 1.05301507573429
  511 1.05301481105847
  };
  \addplot [draw=color0, fill=color0, mark=x, only marks]
  table{%
  x  y
  1 1.05301545871967
  13 1.05301571385372
  26 1.05301598295687
  38 1.05301622248422
  51 1.05301646910065
  64 1.0530167012708
  77 1.05301691880267
  89 1.05301710684271
  102 1.05301729672802
  115 1.05301747185122
  128 1.05301763194903
  141 1.05301777711613
  153 1.05301789799086
  166 1.05301801429211
  179 1.05301811568421
  192 1.05301820186884
  204 1.05301826770028
  217 1.05301832374446
  230 1.05301836424645
  243 1.05301838706854
  255 1.05301839387898
  268 1.05301838619875
  281 1.05301836302064
  294 1.05301832471628
  307 1.05301827142924
  319 1.05301820829722
  332 1.05301812541666
  345 1.05301802812422
  358 1.05301791466654
  370 1.05301779651033
  383 1.05301765433342
  396 1.0530174979105
  409 1.05301732656433
  422 1.05301714083838
  434 1.05301695690823
  447 1.05301674284241
  460 1.05301651470874
  473 1.05301627361249
  485 1.05301603985047
  498 1.05301577510944
  511 1.05301550098737
  };
  \addplot [
    draw=red!54.5098039215686!black,
    fill=red!54.5098039215686!black,
    mark=Custom_8,
    only marks
  ]
  table{%
  x  y
  1 0.80327211516833
  13 0.800100044256811
  26 0.796484589777146
  38 0.792963669311482
  51 0.788926467409445
  64 0.784627692587697
  77 0.780030437461819
  89 0.775482893189413
  102 0.770174746886304
  115 0.76440097803694
  128 0.758070844886965
  141 0.751064224373287
  153 0.74385554244859
  166 0.73502986194581
  179 0.724821946543414
  192 0.712713366242274
  204 0.699099747385768
  217 0.680192045349445
  230 0.653404717462196
  243 0.606971241620078
  255 0.413264493925693
  268 0.601532863461694
  281 0.650802187352946
  294 0.678481445383662
  307 0.697825403523052
  319 0.711681600271431
  332 0.723966370501383
  345 0.734298917042998
  358 0.743217407853847
  370 0.750492942948374
  383 0.757557708153469
  396 0.763935169949832
  409 0.769748210744868
  422 0.775089466457228
  434 0.779663272930444
  447 0.784285247576601
  460 0.788605584610063
  473 0.792661758540986
  485 0.796198274218896
  498 0.799828872639096
  511 0.80327220937088
  };
  \addplot [
    draw=blue!54.5098039215686!black,
    fill=blue!54.5098039215686!black,
    mark=Custom_11,
    only marks
  ]
  table{%
  x  y
  1 1.12280361740398
  13 1.12300759576807
  26 1.12322874407678
  38 1.12343304964316
  51 1.12365457542435
  64 1.12387631827064
  77 1.12409828740199
  89 1.12430339120715
  102 1.12452582420745
  115 1.12474851184882
  128 1.12497145966606
  141 1.12519468110104
  153 1.12540098377177
  166 1.12562476379712
  179 1.12584885176085
  192 1.12607326437862
  204 1.1262807153614
  217 1.12650579746687
  230 1.12673126061917
  243 1.12695713775495
  255 1.12716606219424
  268 1.12697678104278
  281 1.12675081038319
  294 1.12652526516026
  307 1.1263001178474
  319 1.12609262134057
  332 1.1258681681081
  345 1.12564404751768
  358 1.1254202449118
  370 1.12521392803425
  383 1.12499069911739
  396 1.1247677510456
  409 1.12454507234981
  422 1.12432265374411
  434 1.12411756695021
  447 1.12389562073277
  460 1.12367390774113
  473 1.12345241419345
  485 1.1232481417443
  498 1.1230270326783
  511 1.12280610911434
  };
  \addplot [draw=color1, fill=color1, mark=Custom_10, only marks]
  table{%
  x  y
  1 0.819662016772226
  13 0.816575487688
  26 0.813053073782112
  38 0.809618383511294
  51 0.805674978900447
  64 0.801470406213275
  77 0.796967768778757
  89 0.792507942167767
  102 0.787295241614425
  115 0.781617363309002
  128 0.775383574075118
  141 0.768473759446893
  153 0.761354853706747
  166 0.752626890900738
  179 0.742517184528563
  192 0.73050731900322
  204 0.716985291987659
  217 0.698177320376799
  230 0.671490239682042
  243 0.625157676579541
  255 0.431545679608872
  268 0.925163057745125
  281 0.970341334401495
  294 0.996384164431612
  307 1.01466323268606
  319 1.02773486883467
  332 1.03927514228984
  345 1.04892452160272
  358 1.05719654960902
  370 1.06389625959596
  383 1.07035238908507
  396 1.07613212572104
  409 1.08135507781551
  422 1.08611167548208
  434 1.0901492857595
  447 1.09419312801317
  460 1.09793740068925
  473 1.10141897644394
  485 1.1044260027955
  498 1.10748364524483
  511 1.11035444277621
  };
  \coordinate (insetPosition) at (rel axis cs:0.53,0.02);
  \coordinate (tipBig) at (axis cs:255,1.11);
  \draw [gray, thick] (axis cs:200,1.09) rectangle (axis cs:312,1.15);
  \end{axis}
  \input{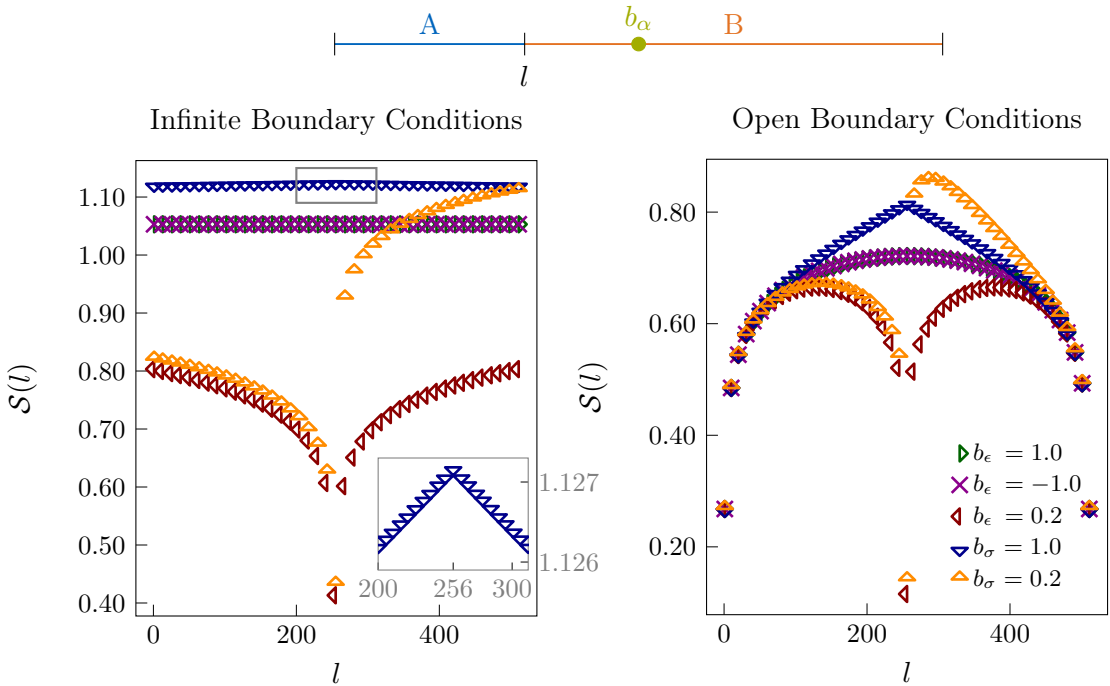}
  \end{tikzpicture}
  
    \end{minipage}
    \begin{minipage}{.45\linewidth}
        \hspace{14 pt} 
\begin{tikzpicture}[
baseline=(current bounding box.center)
]

\definecolor{color0}{rgb}{0.545098039215686,0,0.545098039215686}
\definecolor{color1}{rgb}{1,0.549019607843137,0}

\begin{axis}[
height=3.0in,
legend cell align={left},
legend style={
  fill opacity=0.8,
  draw opacity=1,
  text opacity=1,
  at={(0.97,0.03)},
  anchor=south east,
  draw=none
},
minor xtick={},
minor ytick={},
tick pos=left,
title={Open Boundary Conditions},
width=2.7in,
x grid style={white!69.0196078431373!black},
xlabel={\(\displaystyle l\)},
xmin=-24.5, xmax=536.5,
xtick style={color=black},
xtick={-200,0,200,400,600},
y grid style={white!69.0196078431373!black},
ylabel={\(\displaystyle \mathcal{S}(l)\)},
ymin=0.0784035013765438, ymax=0.89474813710208,
ytick style={color=black},
ytick={0,0.2,0.4,0.6,0.8,1}
]
\addplot [
  draw=green!39.2156862745098!black,
  fill=green!39.2156862745098!black,
  mark=Custom_9,
  only marks
]
table{%
x  y
1 0.267855006710725
10 0.484827635531411
20 0.544412460939672
31 0.581353607808321
41 0.604590810144019
51 0.622485759003084
61 0.636942940649821
72 0.650082629360817
82 0.660167470718884
92 0.668878478460633
102 0.676476809137587
112 0.683149174553851
123 0.689582730070848
133 0.694719576191834
143 0.699260315341226
153 0.703266789184106
164 0.707115256963444
174 0.710148398356188
184 0.71277063469962
194 0.715007135452855
204 0.716878144233318
215 0.718533237058697
225 0.719684964621391
235 0.720510168890954
245 0.721015279465771
256 0.721205742155179
266 0.721048355846915
276 0.720575000738942
286 0.719782051677199
296 0.718663342435159
307 0.717045795804028
317 0.715210387454752
327 0.71301127327929
337 0.710428654453333
347 0.707437897280213
358 0.70364003849803
368 0.699684104997735
378 0.695199339004483
388 0.690125325505611
399 0.683771118374232
409 0.677183119494958
419 0.66968482388174
429 0.661095340366205
439 0.651162552064365
450 0.638243401628117
460 0.624064669255625
470 0.606578606619577
480 0.58400689047305
491 0.548552434013973
501 0.493114665235859
511 0.267855006710771
};
\addlegendentry{$b_{\epsilon \:} = 1.0$}
\addplot [draw=color0, fill=color0, mark=x, only marks]
table{%
x  y
1 0.267855006710733
10 0.484827635530832
20 0.544412460939347
31 0.58135360780894
41 0.604590810144822
51 0.622485759003094
61 0.636942940649954
72 0.650082629357453
82 0.660167470714699
92 0.668878478455377
102 0.676476809105192
112 0.683149174555662
123 0.689582730069046
133 0.694719576191476
143 0.69926031534377
153 0.70326678917639
164 0.707115256960939
174 0.710148398350455
184 0.712770634751233
194 0.715007135457806
204 0.716878144279679
215 0.718533237043387
225 0.719684964606409
235 0.720510168872714
245 0.721015279445852
256 0.721205742054352
266 0.721048355835657
276 0.72057500072404
286 0.71978205166678
296 0.718663342445215
307 0.717045795802219
317 0.715210387467564
327 0.713011273282012
337 0.710428654472663
347 0.707437897300925
358 0.703640038516671
368 0.69968410502233
378 0.695199339018778
388 0.690125325518266
399 0.683771118382663
409 0.677183119510044
419 0.669684823894804
429 0.661095340373118
439 0.651162552071996
450 0.638243401632301
460 0.62406466926184
470 0.606578606623118
480 0.584006890474987
491 0.548552434014548
501 0.493114665236236
511 0.267855006710789
};
\addlegendentry{$b_{\epsilon \:} = -1.0$}
\addplot [
  draw=red!54.5098039215686!black,
  fill=red!54.5098039215686!black,
  mark=Custom_8,
  only marks
]
table{%
x  y
1 0.26785150811734
10 0.484662629848327
20 0.543788639857738
31 0.579881285675501
41 0.602028279369278
51 0.618520774073391
61 0.631252170009022
72 0.64210303657335
82 0.64973270749828
92 0.655608888805417
102 0.659965098285187
112 0.662953978731195
123 0.664775305113838
133 0.665152033738249
143 0.664322727545205
153 0.66226229029111
164 0.658495761418308
174 0.653582912376251
184 0.647062376721026
194 0.638646832290859
204 0.627880822770974
215 0.612396080986165
225 0.593415780538387
235 0.566359925796083
245 0.520835379837247
256 0.115510075727705
266 0.514070499688366
276 0.562945225160797
286 0.591157629215962
296 0.610745414152493
307 0.626649463388659
317 0.637685528917381
327 0.646311995534556
337 0.653006997035373
347 0.658070925081874
358 0.661986434724804
368 0.664172819287383
378 0.665123504039744
388 0.664867277304732
399 0.663181756840683
409 0.660323547378129
419 0.656109857447472
429 0.650394590787468
439 0.642953851204475
450 0.632361752859377
460 0.619941863879843
470 0.603890042088731
480 0.582439340030625
491 0.547866444854027
501 0.492917050392483
511 0.267851508117338
};
\addlegendentry{$b_{\epsilon \:} = 0.2$}
\addplot [
  draw=blue!54.5098039215686!black,
  fill=blue!54.5098039215686!black,
  mark=Custom_11,
  only marks
]
table{%
x  y
1 0.267858755150414
10 0.485004278293206
20 0.545078736866895
31 0.582919473738798
41 0.607301400520369
51 0.626650767697565
61 0.642870055102065
72 0.658300604927858
82 0.67078560898443
92 0.682196127375775
102 0.692790075884042
112 0.702750720687541
123 0.713135088209426
133 0.722162869759825
143 0.730875583642145
153 0.739331121031454
164 0.748388814263265
174 0.756439826001681
184 0.764344514472687
194 0.772123995182014
204 0.77979447416571
215 0.78812044829259
225 0.795597276567867
235 0.802991787800911
245 0.810306557210124
256 0.818260731655133
266 0.811033672473508
276 0.803726832718881
286 0.796340369275432
296 0.788871963112393
307 0.780556025152831
317 0.772895718503879
327 0.765127794054409
337 0.757236488074192
347 0.749201219555452
358 0.74016436265099
368 0.731731876444629
378 0.723047247927132
388 0.714054018814264
399 0.703717450994562
409 0.693812102792589
419 0.683288765142603
429 0.671970010918123
439 0.659606985668391
450 0.644363561264169
460 0.628392078930922
470 0.609420732715845
480 0.585673243451739
491 0.549284880629245
501 0.493326182014364
511 0.267858755150394
};
\addlegendentry{$b_{\sigma} = 1.0$}
\addplot [draw=color1, fill=color1, mark=Custom_10, only marks]
table{%
x  y
1 0.26785184769187
10 0.484678648350473
20 0.543849248833485
31 0.580024569235342
41 0.60227821185072
51 0.618908624506593
61 0.631810886327853
72 0.642890383886451
82 0.650767941911563
92 0.656933843052731
102 0.661626114176226
112 0.665003030888399
123 0.667319647025939
133 0.66821842159869
143 0.667990850050756
153 0.666625430545831
164 0.663751252908812
174 0.659788420820539
184 0.654376670876516
194 0.647260589660832
204 0.638024359291513
215 0.624548235484676
225 0.607760507469543
235 0.583336487190751
245 0.541009346289352
256 0.140209167545688
266 0.828036816984197
276 0.851711958264934
286 0.85764156275092
296 0.856044760849682
307 0.849579039009453
317 0.841066335994184
327 0.8308914091588
337 0.819546312525974
347 0.807351083270971
358 0.793210048105949
368 0.779851255878111
378 0.766110891950981
388 0.752043607317962
399 0.736214297199234
409 0.721481164570466
419 0.706358112488764
429 0.690727894796423
439 0.674395637261537
450 0.655239259814378
460 0.63616977081412
470 0.614577596164743
480 0.588719695620598
491 0.550632033229754
501 0.493716501373243
511 0.267865680484859
};
\addlegendentry{$b_{\sigma} = 0.2$}
\end{axis}

\end{tikzpicture}
    \end{minipage}
    \caption{\label{fig:int_EE}The EE, $\mathcal{S}$, for different bipartionings of the system for infinite (left) and open (right) bcs obtained by the DMRG calculation for energy and duality defects for $\chi = 100$. In the middle of the system is an energy $\epsilon$-defect or a duality $\sigma$-defect of various strength. As for the symmetric EE, the results for the $b_\epsilon=\pm1$ are identical. For the latter, in the infinite system case, the result is a constant EE for different bipartitionings. The actual value of this constant depends on the amount of entanglement truncation~\cite{Tagliacozzo2008, Pollmann2009} and will be different for different values of $\chi$. The corresponding results for the finite system follow the dependence given in footnote~\ref{foot_int_EE}. For defect strengths $b_{\epsilon, \sigma}< 1$, the EE exhibits a dip centered at the defect-bond due to the reflections of the entanglement-carrying modes at the defect. On the other hand, the EE is discontinuous for the duality defect with $b_\sigma< 1$ as the defect is traversed. This is a consequence of the zero energy mode localized at the defect. The situation is different for the topological defect: $b_\sigma= 1$, which is associated with a {\it higher} interface EE compared to $b_\epsilon =1$ (see inset of the left panel for infinite and right panel for finite system results). The scaling of the interface EE with correlation length~(system size) for infinite~(finite) systems is shown in Fig.~\ref{fig:int_EE_scal}.}
\end{figure}

\begin{figure}
    \centering
    \hspace{1.6cm}
\begin{tikzpicture}
    \definecolor{color0}{rgb}{0,0.396078431372549,0.741176470588235}
    \definecolor{color1}{rgb}{0.890196078431372,0.447058823529412,0.133333333333333}
    \definecolor{color2}{rgb}{0.635294117647059,0.67843137254902,0}
    \draw[line width=0.25mm, color0 ]  (0,0) -- (4,0) node[midway,above] {A};
    \draw[line width=0.25mm, color1 ]  (4,0) -- (8,0) node[midway,above] {B};
\node[circle,
    draw = color2,
    fill = color2,
    scale = 0.5] (d) at (4,0) {};
\node[text=color2, below] (d) at (4,0.05) {$b_\alpha$};
    \draw (4,0.15) -- (4, -0.15);
    \draw (0,0.15) -- (0, -0.15);
    \draw (8,0.15) -- (8, -0.15);
\end{tikzpicture}
    \\
    \hspace{-4em}
    \begin{minipage}{.45\linewidth}
\begin{tikzpicture}[
baseline=(current bounding box.center)
]

\definecolor{color0}{rgb}{1,0.549019607843137,0}
\definecolor{color1}{rgb}{1,0.647058823529412,0}
\definecolor{color2}{rgb}{0.545098039215686,0,0.545098039215686}

\begin{axis}[
height=2.5in,
legend cell align={left},
legend style={fill opacity=0.8, draw opacity=1, text opacity=1, at={(0.98,0.55)}, anchor=east, draw=none},
log basis x={10},
minor xtick={20,30,40,50,60,70,80,90,200,300,400,500,600,700,800,900,2000,3000,4000,5000,6000,7000,8000,9000,20000,30000,40000,50000,60000,70000,80000,90000,200000,300000,400000,500000,600000,700000,800000,900000},
minor ytick={},
tick pos=left,
title={Finite Entanglement Scaling},
width=2.7in,
x grid style={white!69.0196078431373!black},
xlabel={\(\displaystyle \xi\)},
xmin=632.146078962473, xmax=2708.9267519698,
xmode=log,
xtick style={color=black},
xtick={700,1000,2000},
xticklabels= {$7\times 10^2$, $10^3$, $2\times 10^3$},
y grid style={white!69.0196078431373!black},
ylabel={\(\displaystyle \mathcal{S}_\mathrm{I} \left(\xi\right)\)},
ymin=0.05, ymax=1.17519381999896,
ytick style={color=black},
ytick={0,0.5,1,1.5}
]
\addplot [
  draw=green!39.2156862745098!black,
  fill=green!39.2156862745098!black,
  mark=Custom_9,
  only marks
]
table{%
x  y
675.373169770727 0.942967116082953
802.329116068855 0.956242145511053
963.875394312973 0.970658078385335
1171.21280326581 0.988711089829465
1460.58528791364 1.0070317994674
1733.79597147754 1.02159054441034
2130.07341838976 1.03821718823977
2535.54257276105 1.05301749781864
};
\addlegendentry{$b_{\epsilon \:} = 1.0$}
\addplot [draw=color2, fill=color2, mark=x, only marks]
table{%
x  y
675.373169770727 0.942989552096631
802.329116068855 0.956238111258386
963.875394312973 0.97064557890083
1171.21280326581 0.988569411703596
1460.58528791364 1.00703100620647
1733.79597147754 1.02158822623221
2130.07341838976 1.03821996852681
2535.54257276105 1.0530183938377
};
\addlegendentry{$b_{\epsilon \:} = -1.0$}
\addplot [
  draw=red!54.5098039215686!black,
  fill=red!54.5098039215686!black,
  mark=Custom_8,
  only marks
]
table{%
x  y
675.373169770727 0.167108532636727
802.329116068855 0.170282263274436
963.875394312973 0.173905128944443
1171.21280326581 0.178095933341822
1460.58528791364 0.182535481316019
1733.79597147754 0.186028853232767
2130.07341838976 0.189894774094601
2535.54257276105 0.193395511606858
};
\addlegendentry{$b_{\epsilon \:} = 0.2$}
\addplot [
  draw=blue!54.5098039215686!black,
  fill=blue!54.5098039215686!black,
  mark=Custom_11,
  only marks
]
table{%
x  y
675.373169770727 1.02061106262586
802.329116068855 1.03422140656844
963.875394312973 1.04860779985076
1171.21280326581 1.06442803029096
1460.58528791364 1.08229775940433
1733.79597147754 1.09609110686264
2130.07341838976 1.11269781554673
2535.54257276105 1.12718349617858
};
\addlegendentry{$b_{\sigma} = 1.0$}
\addplot [draw=color0, fill=color0, mark=Custom_10, only marks]
table{%
x  y
675.373169770727 0.184674941949969
802.329116068855 0.188229169203122
963.875394312973 0.192499913554341
1171.21280326581 0.195967616755528
1460.58528791364 0.200597518049089
1733.79597147754 0.203899351835474
2130.07341838976 0.207974718596061
2535.54257276105 0.211685619232019
};
\addlegendentry{$b_{\sigma} = 0.2$}
\addplot [line width=1pt, green!50.1960784313725!black, opacity=0.7, forget plot]
table {%
675.373169770727 0.942145506547195
802.329116068855 0.956584778159684
963.875394312973 0.971962027849224
1171.21280326581 0.988294045024169
1460.58528791364 1.00680254577354
1733.79597147754 1.02117663952974
2130.07341838976 1.03843159381154
2535.54257276105 1.05303832304986
};
\addplot [line width=1pt, red, opacity=0.7, forget plot]
table {%
675.373169770727 0.166977019771073
802.329116068855 0.170426907992437
963.875394312973 0.174100901667785
1171.21280326581 0.178003012189359
1460.58528791364 0.182425136825836
1733.79597147754 0.18585945249446
2130.07341838976 0.189982074821577
2535.54257276105 0.193471972685145
};
\addplot [line width=1pt, blue, opacity=0.7, forget plot]
table {%
675.373169770727 1.02026572746525
802.329116068855 1.03414025871312
963.875394312973 1.04891608238689
1171.21280326581 1.06460933123738
1460.58528791364 1.08239393830691
1733.79597147754 1.09620584089677
2130.07341838976 1.11278592947661
2535.54257276105 1.12682136884538
};
\addplot [line width=1pt, color1, opacity=0.7, forget plot]
table {%
675.373169770727 0.184864731248951
802.329116068855 0.188353253614994
963.875394312973 0.192068391117068
1171.21280326581 0.196014200069655
1460.58528791364 0.200485846601346
1733.79597147754 0.203958622022838
2130.07341838976 0.208127412218948
2535.54257276105 0.211656392281801
};
\node (c1) at (axis cs: 800, 1.1) {$c\simeq 0.486$};
\node (c2) at (axis cs: 800, 0.82) {$c\simeq 0.504$};
\coordinate (curve1) at (axis cs: 1200, 1.08);
\coordinate (curve2) at (axis cs: 770, 0.94);
\draw[->] (curve1) to[out = 90, in =0] (c1);
\draw[->] (curve2) to[out = -90, in =110] (c2);
\node (offset) at (axis cs: 950, 0.6) {$\simeq \ln(2)/2-1/4$};
\node (ceff1) at (axis cs: 830, 0.27) {$c_\mathrm{eff} \simeq 0.119$};
\node (ceff2) at (axis cs: 1450, 0.1) {$c_\mathrm{eff} \simeq 0.120$};
\coordinate (curve3) at (axis cs: 1200, 0.22);
\coordinate (curve4) at (axis cs: 900, 0.13);
\draw[->] (curve3) to[out = 90, in = 0] (ceff1);
\draw[->] (curve4) to[out = -90, in = 180] (ceff2);
\coordinate (offTop) at (axis cs: 1600,1.08);
\coordinate (offBot) at (axis cs: 1600, 1.02);
\coordinate (offMid) at (axis cs: 1600, 1.05);
\draw[<->] (offTop) -- (offBot);
\draw[-] (offMid) to[out = -180, in = 40] (offset);
\end{axis}
\end{tikzpicture}
    \end{minipage}
    \begin{minipage}{.45\linewidth}
        \hspace{14 pt} 
\begin{tikzpicture}[
baseline=(current bounding box.center)
]

\definecolor{color0}{rgb}{0.545098039215686,0,0.545098039215686}
\definecolor{color1}{rgb}{1,0.549019607843137,0}
\definecolor{color2}{rgb}{1,0.647058823529412,0}

\begin{axis}[
height=2.5in,
log basis x={2},
minor xtick={},
minor ytick={},
tick pos=left,
title={Finite Size Scaling},
width=2.7in,
x grid style={white!69.0196078431373!black},
xlabel={\(\displaystyle L\)},
xmin=461.440236856745, xmax=4544.79655758989,
xmode=log,
xtick style={color=black},
xtick={128,256,512,1024,2048,4096,8192,16384},
y grid style={white!69.0196078431373!black},
ylabel={\(\displaystyle \mathcal{S}_\mathrm{I} \left(L\right)\)},
ymin=0.0, ymax=1.03510212721308,
ytick style={color=black},
ytick={0,0.25,0.5,0.75,1,1.25}
]
\addplot [
  draw=green!39.2156862745098!black,
  fill=green!39.2156862745098!black,
  mark=Custom_9,
  only marks
]
table{%
x  y
512 0.721205742054352
724 0.750166246014871
1024 0.779118695048479
1448 0.80803458617324
2048 0.83695321786261
2896 0.865835644343667
4096 0.894694446783879
};
\addplot [draw=color0, fill=color0, mark=x, only marks]
table{%
x  y
512 0.721205742155179
724 0.75016624599776
1024 0.779118695397928
1448 0.808034587323568
2048 0.836953218403226
2896 0.865835642941876
4096 0.894694454780875
};
\addplot [
  draw=red!54.5098039215686!black,
  fill=red!54.5098039215686!black,
  mark=Custom_8,
  only marks
]
table{%
x  y
512 0.115510075724411
724 0.12198391388413
1024 0.128463385873671
1448 0.1349399022198
2048 0.141420403012679
2896 0.147893162818133
4096 0.154351866077884
};
\addplot [
  draw=blue!54.5098039215686!black,
  fill=blue!54.5098039215686!black,
  mark=Custom_11,
  only marks
]
table{%
x  y
512 0.818260731834834
724 0.847080218032528
1024 0.875932801304528
1448 0.904777626642332
2048 0.933643872714747
2896 0.962481678092652
4096 0.991286696615852
};
\addplot [draw=color1, fill=color1, mark=Custom_10, only marks]
table{%
x  y
512 0.140209167545688
724 0.146663911690234
1024 0.153129844353574
1448 0.159596472420173
2048 0.166068320450583
2896 0.172528274851157
4096 0.178951727258851
};
\addplot [line width=1pt, green!50.1960784313725!black, opacity=0.7]
table {%
512 0.721257964548778
724 0.750164745515078
1024 0.779089351608862
1448 0.807996132575161
2048 0.836920738668945
2896 0.865827519635245
4096 0.894752125729029
};
\addplot [line width=1pt, red, opacity=0.7]
table {%
512 0.11551329059738
724 0.121986326384668
1024 0.128463353715956
1448 0.134936389503244
2048 0.141413416834532
2896 0.14788645262182
4096 0.154363479953107
};
\addplot [line width=1pt, blue, opacity=0.7]
table {%
512 0.818256626279834
724 0.847090303462732
1024 0.875941760694272
1448 0.90477543787717
2048 0.933626895108709
2896 0.962460572291608
4096 0.991312029523147
};
\addplot [line width=1pt, color2, opacity=0.7]
table {%
512 0.140211790665221
724 0.14667033068741
1024 0.153132853314904
1448 0.159591393337092
2048 0.166053915964587
2896 0.172512455986775
4096 0.17897497861427
};
\coordinate (topLine) at (axis cs: 1500, 0.92);
\coordinate (botLine) at (axis cs: 1800, 0.81);
\node (c1) at (axis cs: 740, 0.955) {$c \simeq 0.500$};
\node (c2) at (axis cs: 3200, 0.72) {$c \simeq 0.501$};
\draw[->] (topLine) to[out = 90, in = 0] (c1);
\draw[->] (botLine) to[out = -90, in = 180] (c2); 
\node (offset) at (axis cs: 1000, 0.5) {$\simeq \ln(2)/2-1/4$};
\node (ceff1) at (axis cs: 2046, 0.25) {$c_\mathrm{eff} \simeq 0.112$};
\node (ceff2) at (axis cs: 1030, 0.05) {$c_\mathrm{eff} \simeq 0.112$};
\coordinate (offTop) at (axis cs: 900,0.85);
\coordinate (offBot) at (axis cs: 900,0.79);
\coordinate (offMid) at (axis cs: 900, 0.82);
\draw[<->] (offTop) -- (offBot);
\draw[-] (offMid) to[out = -160, in = 90] (offset.150);
\coordinate (topLine2) at (axis cs: 1000, 0.2);
\coordinate (botLine2) at (axis cs: 2500, 0.12);
\draw[<-] (ceff1) to[out = 180, in = 90] (topLine2);
\draw[<-] (ceff2)  to[out = 0, in = -90] (botLine2);
\end{axis}
\end{tikzpicture}
    \end{minipage}
    \caption{\label{fig:int_EE_scal}Scaling of the interface EE, $S_\mathrm{I}$, with correlation length~($\xi$) and system size~($L$) for infinite~(left panel) and open~(right panel) bcs. For the infinite bc, the bond dimension~($\chi$) was increased from $50$ to $100$ to increase the correlation length. For both bcs, the $b_\epsilon=\pm1$ and $b_\sigma = 1$, the coefficient of the logarithmic scaling~[see Eq.~\eqref{eq:S_int_1}, with $L\rightarrow\xi$ for infinite bc] gives a central charge of $c \simeq 0.5$. The same for $b_{\epsilon, \sigma} = 0.2$ leads to an effective central charge $c_\mathrm{eff}\simeq0.112$, in agreement with see Eq.~\eqref{ceff}. The fit errors obtained are in the fourth decimal places. We notice that the infinite system scaling results are slightly worse than the finite system. We believe this is due to the fact that the correlation length computed for the translation-invariant system is close, but not the same as the system with defects. Fitting to Eq.~\eqref{eq:S_int_1}, with $L\rightarrow\xi$ for infinite bc, we compute the offsets, $S''_0$ for the different defects. The full dependence of the offsets on the defects is shown in Fig.~\ref{ceff_var}. In particular, the obtained difference between the $b_\sigma = 1.0$ and $b_\epsilon = 1.0$ cases equals $\ln(2)/2-1/4$ up to second digit of precision. This is in agreement with Eq.~\eqref{eq:DeltaInterfaceEntr} for both infinite and open bcs and compatible with the result obtained for periodic bcs in Ref.~\cite{Roy2021a}.}
\end{figure}
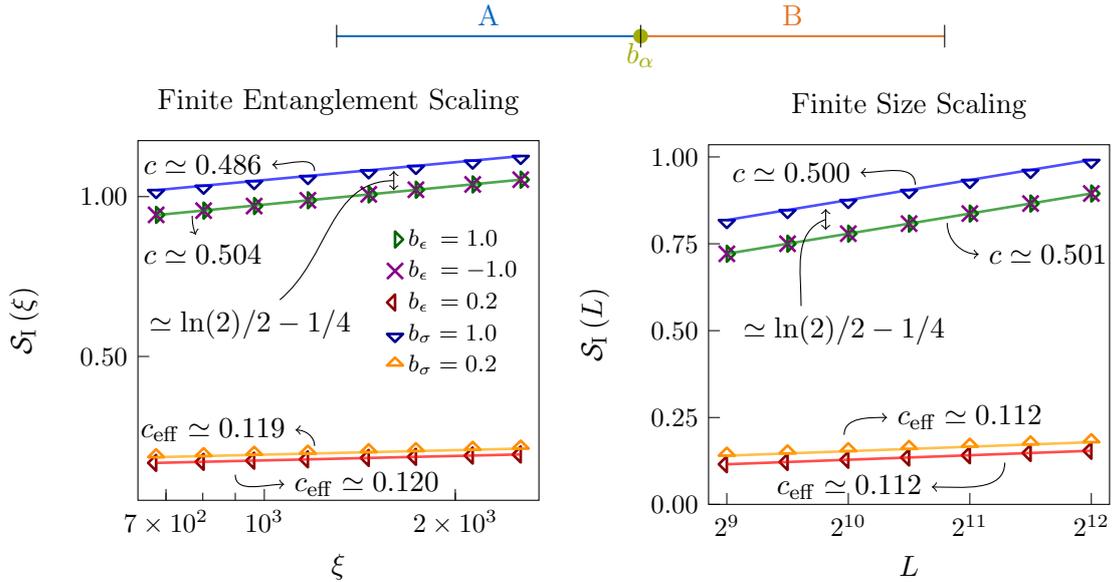

First, consider the energy defect. For $b_\epsilon = \pm1$, the EEs are identical and constant for different bipartitionings for an infinite system. That they are identical and constant is because the system is translation-invariant~(recall that the Hamiltonian for $b_\epsilon = -1$ can be reduced to that for $b_\epsilon=1$ by a unitary transformation). For $b_\epsilon  = 0.2$, the EE is similar to the case without defects sufficiently far away from the defect~(left panel of Fig.~\ref{fig:int_EE}). However, as the defect is approached, the EE dips below the value obtained without defects. This entanglement dip is due to the reflections of entanglement carrying modes at the defect. For finite systems, similar results arise with the important distinction that in absence of the defect, the EE is not a constant value, but rather obeys a logarithmic dependence with subsystem size~(see footnote~\ref{foot_int_EE}). As for the infinite case, introduction of a defect leads to a dip in the EE for partitionings of the system close to the defect bond. In the presence of defects, the explicit form of the dependence of the EE for arbitrary bipartionings is not known. However, the interface EE~(when the bipartioning is done by cutting the system at the defect bond) exhibits the predicted logarithmic scaling with the total system size~(correlation length) for finite~(infinite) systems~[see Eq.~\eqref{eq:S_int_2} and Fig.~\ref{fig:int_EE_scal}]. The central charge obtained is close to the predicted value~\cite{Eisler2010, Peschel2012e, Calabrese2011ru, Brehm2015}:
\begin{align}
\label{ceff}
c_{\rm eff}(s) = \frac{s}{6} - \frac{1}{6} - \frac{1}{\pi^2}\Big[(s+1)\ln(s+1)\ln s + (s-1){\rm Li}_2(1-s) + (s+1){\rm Li}_2(-s)\Big],
\end{align}
where $s = |\sin(2\phi_0)|$ and ${\rm Li}_2$ is the dilogarithm function~\cite{Lewin1981} and $b_\epsilon = \cot\phi_0$. The complete dependence of the effective central charge and the analytical predictions are shown in the left panel of Fig.~\ref{ceff_var}. The offset,~$S''_0$, for different defect strengths is plotted in the top right panel of Fig.~\ref{ceff_var}. We are not aware of any analytical predictions for the universal part, if any, of this offset for the energy defect. 

Now, consider the duality defect away from the topological point~($b_\sigma\neq1$). As expected, sufficiently far away from the defect, the EEs approach the case without defects. As the defect location is approached from the left, the EE goes down as in the energy defect. However, upon crossing the duality defect, the EE exhibits a discontinuous jump~(see Fig.~\ref{fig:int_EE}). This is due to the localized zero mode that is included in the subsystem~(see Fig.~\ref{def_schematic} and surrounding discussions). Despite this discontinuous behavior, for partioning at the defect bond, the interface EE exhibits a logarithmic scaling with an effective central charge which is identical to the energy defect case~[Eq.~\eqref{ceff}]. We see that the offset computed for the duality defect is {\it different} from that obtained for the energy defect for a given defect strength. However, we are not aware of an analytical prediction for the offset for $b_\sigma\neq1$. Only for the topological case are both the effective central charge and the offset analytically predictable~\cite{Roy2021a}. This is discussed next.

The topological point corresponds to the case when the interface EE satisfies Eq.~\eqref{eq:S_int_2} with $c_{\rm eff} = c=1/2$~(see Sec.~\ref{sec:EE_scaling}), which is indicative of the perfect transmission of information across the defect. The results for this case are shown in Fig.~\ref{fig:int_EE}. As the non-topological case, the EE for partitioning the system far from the defect is close to the results obtained for without any defect. However, as the subsystem size increases, the deviation from the no-defect case becomes apparent. Unlike the non-topological case, the dip in the EE is replaced by a peak that is {\it higher} than the EE curve for the case without defects. That there is no dip is due to the perfect transmission of information. That the interface EE is even higher than the no-defect case is due to the presence of the zero energy modes in the system. The scaling of the interface EE with the correlation length~(system size) for infinite~(open) bc reveals that the offset $S''_0$ is larger than that obtained for the no-defect case by: 
\begin{equation}
\label{eq:DeltaInterfaceEntr}
\Delta S''_0 = S''_0(b_\sigma = 1) - S''_0(b_\epsilon = 1) = \frac{1}{2}\Delta S\Big(\frac{1}{2}\Big) = -\frac{1}{4} + \frac{1}{2}\ln2,
\end{equation}
where $\Delta S$ is defined in Eq.~\eqref{eq:dS_K}. As pointed out in Ref.~\cite{Roy2021a}, this offset occurs entirely due to a ‘finite-size effect’ correction arising due to the existence of nonlocal zero-modes. This finite size effect is unavoidable since the interface EE here is computed between left and right halves of the system, {\it i.e.}, between subsystem and the rest each being of size $L/2$. Importantly, this {\it positive} offset bears no relationship to the specific modular S-matrix elements and the claimed offset of $-\ln2$ predicted in Refs.~\cite{Gutperle2015, Brehm2015}. The source of the discrepancy can be traced back to the fact that the twisted torus partition functions do not faithfully capture the geometric setup of the interface EE computation. It is worth emphasizing that this discrepancy plagues {\it only} the subleading/boundary term of the interface EE and not the logarithmic/bulk scaling for all values of $b_\sigma$. This is why the effective central charge is consistent with the field theory computations~[see Eq.~\eqref{ceff}], but not the subleading term. For more details regarding subtle finite size effects in the topological case, we refer the reader to Ref.~\cite{Roy2021a}. 

To summarize, in this section, we have computed the symmetric and interface EEs for the Ising CFT with energy and duality defects. The two computed EEs quantify the entanglement between the subsystem and the rest when the defect is located symmetrically within and at the edge of the subsystem respectively. However, the two computed EEs are measures of entanglement only when the total system is in a pure state. In the next section, we analyze the role of defects in entanglement between subsystems when they together are in a mixed state.  

\section{Entanglement Negativity in the Ising CFT with defects}
\label{sec:EN}
Here, we compute the entanglement between two disjoint subsystems, $B_1$ and $B_2$, separated by a segment of size $l$, with the defect located symmetrically within A~[see Fig.~\ref{fig:geoms}(c)]. In this case, the von-Neumann entropy does not quantify the entanglement between $B_1$ and $B_2$ since the state of $B_1\cup B_2$ is mixed, even though $A\cup B_1\cup B_2$ is in a pure state~\cite{Amico2008, Horodecki_Horodecki_2009}. Measures quantifying mixed state entanglement are nontrivial to compute for many-body systems. A well-known exception is the log-EN~\cite{Vidal2002_EN}~[see Eq.~\eqref{eq:EN}]. The latter has been computed analytically for CFTs using the replica technique~\cite{Calabrese2012, Calabrese2013_EN, Calabrese:2013mi} and has been shown to reveal certain characteristics of the CFTs. The EN computations have since been generalized to other systems: massive deformations of CFTs~\cite{Blondeau_Fournier_2016}, non-equilibrium systems~\cite{Hoogeveen:2014bqa, Eisler_2014} and  Kondo spin chains~\cite{Bayat2010, Bayat2012, Bedoor2016} to name a few. 

As we show here, the log-EN finds a natural application in quantifying entanglement between the subsystems in the configuration described in Fig.~\ref{fig:geoms}(c). Thus far, to the best of our knowledge, the log-EN has never been computed for defect CFTs analytically or numerically. Here, we compute the log-EN using DMRG. As shown in Appendix~\ref{sec:App_DMRG}, the computation of the log-EN is as resource-consuming as that of the symmetric EE. Yet, as shown below, in the presence of defects, the log-EN shows dramatically different results compared to the symmetric EE. Before presenting our numerical results, we summarize the expected scaling behavior of the log-EN. 

\subsection{Scaling of Entanglement Negativity}
\label{sec:EN_scaling}
Consider first the case without any defect. It is known that for certain configurations of subsystems, the log-EN exhibits logarithmic scaling behavior similar to the EE. In fact, for $A=\emptyset$ in Fig.~\ref{fig:geoms}(c), the log-EN reduces to the Renyi entropy with Renyi index $n = 1/2$~\cite{Calabrese2012}. The latter exhibits the characteristic logarithmic scaling of EEs with the system size. The situation is more complex when $A$ has a finite extent. In particular, it can be shown that without a defect, for two disjoint semi-infinite intervals, $B_1, B_2$, the log-EN scales as
\begin{equation}
\label{eq:log_EN}
{\cal E}(l) = -\frac{c}{4}\ln l + {\cal E}_0,
\end{equation}
where $l$, the size of the segment A, is the separation between $B_1$ and $B_2$ and ${\cal E}_0$ is a non-universal constant.  Note the {\it negative} sign in front of the logarithm. It is useful to contrast this setup with the symmetric EE setup of Fig.~\ref{fig:geoms}(a). The symmetric EE scales also logarithmically with the size of the segment A, but with a coefficient $+c/3$~[see Eq.~\eqref{eq:S_block}].  Importantly, the symmetric EE quantifies the entanglement between A and $B_1\cup B_2$ and {\it not} between $B_1$ and $B_2$. 

Now, consider the case when there are defects. Unlike the symmetric EE, the log-EN contains signatures of the defect both in the leading logarithmic scaling and the subleading terms. As we numerically demonstrate below, the defects, being marginal perturbations, lead to a continuously varying central charge in the scaling of the log-EN:
\begin{equation}
\label{eq:log_EN_eff}
{\cal E}(l) = -\frac{\tilde{c}_{\rm eff}}{4}\ln l + \tilde{\cal E}_0,
\end{equation}
where $\tilde{c}_{\rm eff}$ depends on the defect strength~[compare Eq.~\eqref{eq:S_block_1}]. In particular, for perfectly reflective~($b_{\epsilon, \sigma} = 0$) and  perfectly transmissive~($b_{\epsilon, \sigma} = 1$) defects, $\tilde{c}_{\rm eff} = 0$ and $1/2$ respectively. This is indicative of the expected zero and maximal entanglement between the blocks $B_1$ and $B_2$. In between these two points, $\tilde{c}_{\rm eff}$ grows monotonically. We emphasize that except for the two special points, $\tilde{c}_{\rm eff}\neq c_{\rm eff}$, the latter being the effective central charge obtained for the interface EE~[see Eq.~\eqref{ceff}]. This difference can be traced to the partial transposition in the definition of the EN. The subleading term, $ \tilde{\cal E}_0$, also contains information about the defects, similar to $\tilde{S}_0$ and $S''_0$ for the symmetric and interface EEs. Note that Eq.~\eqref{eq:log_EN} is valid only when the two blocks $B_1$ and $B_2$ are both semi-infinite. For finite-size systems, we find an analogous scaling with the usual substitution of $l$ by $L\sin(\pi l/L)/\pi$, where $L$ is the total system-size.

We are not aware of any analytical computation of $\tilde{c}_{\rm eff}$ and $\tilde{\cal E}_0$ in the presence of defects. In the following section, we present numerical results obtained using DMRG. As shown below, the zero energy modes associated with the duality defect~(see Sec.~\ref{sec:defects}) also lead to nontrivial finite-size corrections that lead to deviation from the logarithmic scaling given in Eq.~\eqref{eq:log_EN_eff} for the log-EN. We exhibit this explicitly by performing the DMRG simulations with infinite and open bcs. 

\subsection{DMRG results}
\label{sec:EN_res}
\begin{figure}
    \centering
    \hspace{1.8cm}
\begin{tikzpicture}
    \definecolor{color0}{rgb}{0,0.396078431372549,0.741176470588235}
    \definecolor{color1}{rgb}{0.890196078431372,0.447058823529412,0.133333333333333}
    \definecolor{color2}{rgb}{0.635294117647059,0.67843137254902,0}
    \definecolor{color3}{rgb}{0,0.2,0.349019607843137}

    \draw
        (0,0.15) -- (0, -0.15);
    \draw
        [line width = 0.25mm, color1]
        (0,0)--(3,0)
        node[midway, above]{$B_1$};
    \draw
        [line width = 0.25mm, color0]
        (3,0) -- (5,0)
        node[pos=0.25,above, color0] {A}
        node[pos = 0.75,below,black] {$l$};
    \draw
        [line width = 0.25mm, color1] (5,0) -- (8,0)
        node[midway, above]{$B_2$};
    \draw
        (8,0.15) -- (8, -0.15);
    \node[circle,
    		draw = color2,
    		fill = color2,
    		scale = 0.5] (d) at (4,0) {};
\node[text=color2, above] (b) at (4,0.05) {$b_\alpha$};
    \draw
        (3,0.15) -- (3,-0.15);
    \draw
        (5,0.15) -- (5,-0.15);    
\end{tikzpicture}
    \\
    \hspace{-4em}
    \begin{minipage}{.45\linewidth}
\begin{tikzpicture}[
baseline=(current bounding box.center)
]

\definecolor{color0}{rgb}{0.545098039215686,0,0.545098039215686}
\definecolor{color1}{rgb}{1,0.549019607843137,0}
\definecolor{color2}{rgb}{1,0.647058823529412,0}

\begin{axis}[
height=3.0in,
log basis x={10},
minor xtick={0.2,0.3,0.4,0.5,0.6,0.7,0.8,0.9,2,3,4,5,6,7,8,9,20,30,40,50,60,70,80,90,200,300,400,500,600,700,800,900,2000,3000,4000,5000,6000,7000,8000,9000,20000,30000,40000,50000,60000,70000,80000,90000},
minor ytick={},
tick pos=left,
title={Infinite Boundary Conditions},
width=2.7in,
x grid style={white!69.0196078431373!black},
xlabel={\(\displaystyle l\)},
xmin=1.56978367970064, xmax=323.611467343625,
xmode=log,
xtick style={color=black},
xtick={0.1,1,10,100,1000,10000},
y grid style={white!69.0196078431373!black},
ylabel={\(\displaystyle \mathcal{E}(l)\)},
ymin=-0.05, ymax=1.2,
ytick style={color=black},
ytick={-0.2,0,0.2,0.4,0.6,0.8,1,1.2}
]
\addplot [
  draw=green!39.2156862745098!black,
  fill=green!39.2156862745098!black,
  mark=Custom_9,
  only marks
]
table{%
x  y
2 1.05649649282041
4 0.95294047560352
6 0.898450472151941
8 0.861113416664006
10 0.832584520909396
12 0.80945440825487
16 0.773167052824328
20 0.745129919489808
26 0.712237649611951
34 0.67865439313647
42 0.652221929946376
56 0.616256140349733
70 0.5883648726071
92 0.554207972856668
118 0.523131586912219
152 0.491565953006433
196 0.45986538859967
254 0.427486676374094
};
\addplot [draw=color0, fill=color0, mark=x, only marks]
table{%
x  y
2 1.05649767605981
4 0.952941732991871
6 0.898451732561783
8 0.861114577473588
10 0.832585602292251
12 0.809455499553031
16 0.773168194328977
20 0.745130708779822
26 0.712238204827564
34 0.678655078818957
42 0.652222851636518
56 0.616257040415271
70 0.588365971448105
92 0.554209695482909
118 0.523133945542226
152 0.491568422787271
196 0.459867595507855
254 0.427488802964585
};
\addplot [
  draw=red!54.5098039215686!black,
  fill=red!54.5098039215686!black,
  mark=Custom_8,
  only marks
]
table{%
x  y
2 0.419801655421666
4 0.387487261532206
6 0.368136109864713
8 0.353977855611445
10 0.342744646099939
12 0.333417631937057
16 0.318465416627525
20 0.306703884312118
26 0.29273250507129
34 0.278327389645072
42 0.266915084261194
56 0.251311654109571
70 0.239173788966512
92 0.22426916964596
118 0.210674397805039
152 0.196834804286644
196 0.182920283832595
254 0.168714143249253
};
\addplot [
  draw=blue!54.5098039215686!black,
  fill=blue!54.5098039215686!black,
  mark=Custom_11,
  only marks
]
table{%
x  y
2 0.913525166190505
4 0.811447832714341
6 0.755501886571608
8 0.716912692165193
10 0.687446100892565
12 0.66360931866322
16 0.626343714176334
20 0.597669754084006
26 0.564158633151801
34 0.53007573446798
42 0.503337701094654
56 0.467070339730921
70 0.439029067159484
92 0.40475951619409
118 0.373592566570282
152 0.341932361649458
196 0.310243518104739
254 0.27815050813048
};
\addplot [draw=color1, fill=color1, mark=Custom_10, only marks]
table{%
x  y
2 0.367107403868444
4 0.329206322485181
6 0.306733691214292
8 0.29081661167706
10 0.27849604639327
12 0.268445466399918
16 0.252612800261264
20 0.240353090906694
26 0.225960889097719
34 0.211272349284357
42 0.199725910155352
56 0.184040971863729
70 0.171913556230023
92 0.157124681589308
118 0.143716621613832
152 0.130153885067958
196 0.116683825940589
254 0.103147695539319
};
\addplot [line width=1pt, green!50.1960784313725!black, opacity=0.7]
table {%
2 1.03346275234967
4 0.946727252365333
6 0.895990237393196
8 0.859991752380998
10 0.832069158111938
12 0.809254737408861
16 0.773256252396663
20 0.745333658127603
26 0.712503263234809
34 0.678934619146429
42 0.652492904595047
56 0.616494419582848
70 0.588571825313788
92 0.554373844396612
118 0.523228766233896
152 0.491545635605256
196 0.459732586769034
254 0.427295691809619
};
\addplot [line width=1pt, red, opacity=0.7]
table {%
2 0.430933044197848
4 0.393469032481632
6 0.371553990501068
8 0.356005020765416
10 0.343944302846777
12 0.334089978784852
16 0.318541009049201
20 0.306480291130562
26 0.29229972724226
34 0.277800288423649
42 0.266379212809526
56 0.250830243073875
70 0.238769525155235
92 0.223998254360885
118 0.210545636074241
152 0.196860614632131
196 0.183119477098549
254 0.169108879211678
};
\addplot [line width=1pt, blue, opacity=0.7]
table {%
2 0.888368885266505
4 0.800916443113186
6 0.749760043857008
8 0.713464000959866
10 0.685310602864201
12 0.662307601703688
16 0.626011558806547
20 0.597858160710882
26 0.564756394873926
34 0.530910277588218
42 0.504249999931929
56 0.467953957034788
70 0.439800558939122
92 0.405319902824154
118 0.373917384283563
152 0.341972365818948
196 0.309896355263029
254 0.277191341968774
};
\addplot [line width=1pt, color2, opacity=0.7]
table {%
2 0.364813985658846
4 0.327278874246337
6 0.305322241609629
8 0.289743762833829
10 0.277660155925415
12 0.267787130197121
16 0.252208651421321
20 0.240125044512907
26 0.225917568563149
34 0.211390612518026
42 0.199947861835744
56 0.184369383059943
70 0.17228577615153
92 0.157486472244353
118 0.144008323401031
152 0.130297330344465
196 0.116530114698566
254 0.102492927313962
};
\coordinate (topLine1) at (axis cs: 10, 0.9);
\coordinate (botLine1) at (axis cs: 4, 0.77);
\node (c1) at (axis cs:60, 0.85) {$c \simeq 0.501$};
\node (c2) at (axis cs:4, 0.65) {$c \simeq 0.505$};
\draw[<-] (c1) to[out = 180, in = 60] (topLine1);
\draw[<-] (c2)  to[out = 90, in = -90] (botLine1);
\coordinate (offsetTop1) at (axis cs: 3, 0.975);
\coordinate (offsetBot1) at (axis cs: 3, 0.85);
\coordinate (offsetMid1) at (axis cs: 3, 0.9);
\draw[<->] (offsetTop1) -- (offsetBot1);
\node (offsetLabel1) at (axis cs:10, 1.05) {$ \simeq 0.144$};
\draw[-] (offsetLabel1) to[out = 180, in = 0] (offsetMid1);

\coordinate (topLine2) at (axis cs: 5, 0.37);
\coordinate (botLine2) at (axis cs: 80, 0.15);
\node (sig1) at (axis cs:5,0.15) {$\tilde{c}_\mathrm{eff} \simeq 0.216$};
\node (sig2) at (axis cs:40,0.05) {$\tilde{c}_\mathrm{eff} \simeq 0.217$};
\draw[<-] (sig1) to[out = 90, in = -90] (topLine2);
\draw[<-] (sig2)  to[out = 80, in = -90] (botLine2);
 \coordinate (offsetTop2) at (axis cs: 3, 0.4);

\end{axis}

\end{tikzpicture}
    \end{minipage}
    \begin{minipage}{.45\linewidth}
        \vspace{12 pt}
        \hspace{14 pt}
\begin{tikzpicture}[
baseline=(current bounding box.center)
]

\definecolor{color0}{rgb}{0.545098039215686,0,0.545098039215686}
\definecolor{color1}{rgb}{1,0.549019607843137,0}
\definecolor{color2}{rgb}{1,0.647058823529412,0}

\begin{axis}[
height=3.0in,
legend cell align={left},
legend style={fill opacity=0.8, draw opacity=1, text opacity=1, draw=none},
log basis x={10},
minor xtick={0.2,0.3,0.4,0.5,0.6,0.7,0.8,0.9,2,3,4,5,6,7,8,9,20,30,40,50,60,70,80,90,200,300,400,500,600,700,800,900,2000,3000,4000,5000,6000,7000,8000,9000,20000,30000,40000,50000,60000,70000,80000,90000},
minor ytick={},
tick pos=left,
title={Open Boundary Conditions},
width=2.7in,
x grid style={white!69.0196078431373!black},
xlabel={\(\displaystyle \dfrac{L}{\pi}\sin \left(\pi\dfrac{l}{L} \right)\)},
xmin=1.6049654226969, xmax=203.067676664786,
xmode=log,
xtick style={color=black},
xtick={0.1,1,10,100,1000,10000},
y grid style={white!69.0196078431373!black},
ylabel={\(\displaystyle \mathcal{E}(l)\)},
ymin=-0.0289979229727497, ymax=0.609152815231203,
ytick style={color=black},
ytick={-0.1,0,0.1,0.2,0.3,0.4,0.5,0.6,0.7}
]
\addplot [
  draw=green!39.2156862745098!black,
  fill=green!39.2156862745098!black,
  mark=Custom_9,
  only marks
]
table{%
x  y
1.9999498009741 0.58014596349466
3.99959841686457 0.476591615230905
5.99864470794513 0.422104963795172
7.99678762519743 0.384786077436301
9.99372625564787 0.356281725924835
11.989159867684 0.333173346135729
15.9743102885699 0.296921782459357
19.9498383822559 0.268926920307283
25.8898521722725 0.236144347461962
31.7947792366146 0.210336098921163
41.5366446596453 0.176829388038372
51.1221755311639 0.150905705356296
64.2107194517257 0.122626500887426
82.0639977538064 0.0924271641172575
101.83626370635 0.0659984718501549
129.70125031778 0.0363191068281718
162.962390033807 0.00428541123983598
};
\addlegendentry{$b_{\epsilon \:} = 1.0$}
\addplot [draw=color0, fill=color0, mark=x, only marks]
table{%
x  y
1.9999498009741 0.580145959107216
3.99959841686457 0.476591610552413
5.99864470794513 0.422104958949853
7.99678762519743 0.384786072799431
9.99372625564787 0.356281721089357
11.989159867684 0.333173340636193
15.9743102885699 0.296921775881229
19.9498383822559 0.268926914599736
25.8898521722725 0.236144345649934
31.7947792366146 0.210336097398902
41.5366446596453 0.176829382004326
51.1221755311639 0.150905701016231
64.2107194517257 0.122626499157458
82.0639977538064 0.0924271609969276
101.83626370635 0.0659984707302265
129.70125031778 0.0363191124065331
162.962390033807 0.00428540580676709
};
\addlegendentry{$b_{\epsilon \:} = -1.0$}
\addplot [
  draw=red!54.5098039215686!black,
  fill=red!54.5098039215686!black,
  mark=Custom_8,
  only marks
]
table{%
x  y
1.9999498009741 0.201599232002962
3.99959841686457 0.169299185684662
5.99864470794513 0.150017775254177
7.99678762519743 0.135985008886275
9.99372625564787 0.124917013466297
11.989159867684 0.115783753659682
15.9743102885699 0.10128360004219
19.9498383822559 0.0900274291756993
25.8898521722725 0.0768930433919085
31.7947792366146 0.066662337323675
41.5366446596453 0.0535681224498812
51.1221755311639 0.0435933647244911
64.2107194517257 0.0329423953408428
82.0639977538064 0.0221319138400369
101.83626370635 0.0136303680769787
129.70125031778 0.00582174588460659
162.962390033807 0.000260615154913209
};
\addlegendentry{$b_{\epsilon \:} = 0.2$}
\addplot [
  draw=blue!54.5098039215686!black,
  fill=blue!54.5098039215686!black,
  mark=Custom_11,
  only marks
]
table{%
x  y
1.9999498009741 0.467080496581034
3.99959841686457 0.365328940037382
5.99864470794513 0.309670883236211
7.99678762519743 0.271567713844468
9.99372625564787 0.242747699778233
11.989159867684 0.219674210745471
15.9743102885699 0.184216745782334
19.9498383822559 0.157662448558699
25.8898521722725 0.127776575505212
31.7947792366146 0.105406597327745
41.5366446596453 0.0783017336909514
51.1221755311639 0.0591756734274661
64.2107194517257 0.0405572520624219
82.0639977538064 0.0238250145093554
101.83626370635 0.0124273049184839
129.70125031778 0.00390940593306311
162.962390033807 6.79066681013932e-05
};
\addlegendentry{$b_{\sigma} = 1.0$}
\addplot [draw=color1, fill=color1, mark=Custom_10, only marks]
table{%
x  y
1.9999498009741 0.170508665712301
3.99959841686457 0.133704632406295
5.99864470794513 0.112328208254331
7.99678762519743 0.0974807727004549
9.99372625564787 0.0862263814996052
11.989159867684 0.0772448036039311
15.9743102885699 0.0635600970142607
19.9498383822559 0.0534580029135279
25.8898521722725 0.0423040943916305
31.7947792366146 0.0341449859376561
41.5366446596453 0.0245305554667644
51.1221755311639 0.0179738891131092
64.2107194517257 0.0118365870306768
82.0639977538064 0.00660116244097097
101.83626370635 0.00323063872523671
129.70125031778 0.000887257477334841
162.962390033807 8.92876379357599e-06
};
\addlegendentry{$b_{\sigma} = 0.2$}
\addplot [line width=1pt, green!50.1960784313725!black, opacity=0.7, forget plot]
table {%
1.9999498009741 0.559089878856868
3.99959841686457 0.471650691510283
5.99864470794513 0.420512322761617
7.99678762519743 0.384240007350382
9.99372625564787 0.356116323474537
11.989159867684 0.333149153455976
15.9743102885699 0.29694337889334
19.9498383822559 0.268905281942102
25.8898521722725 0.236023729941101
31.7947792366146 0.210103662289975
41.5366446596453 0.176383904194306
51.1221755311639 0.150187377260521
64.2107194517257 0.121428570669314
82.0639977538064 0.0904773955656715
101.83626370635 0.0632432462924135
129.70125031778 0.0327288144332969
162.962390033807 0.0039278815463113
};
\addplot [line width=1pt, red, opacity=0.7, forget plot]
table {%
1.9999498009741 0.206389232655521
3.99959841686457 0.171328684159154
5.99864470794513 0.150823698786508
7.99678762519743 0.13627956460431
9.99372625564787 0.125002792571683
11.989159867684 0.115793631294048
15.9743102885699 0.101276178039704
19.9498383822559 0.0900337238525639
25.8898521722725 0.0768491863065167
31.7947792366146 0.0664559997217728
41.5366446596453 0.0529353661263011
51.1221755311639 0.0424313274810445
64.2107194517257 0.0308998898752484
82.0639977538064 0.0184893768241661
101.83626370635 0.00756928206656468
};
\addplot [line width=1pt, blue, opacity=0.7, forget plot]
table {%
1.9999498009741 0.437579017476751
3.99959841686457 0.353154118060339
5.99864470794513 0.303778640440531
7.99678762519743 0.268756739241291
9.99372625564787 0.241602562098428
11.989159867684 0.219427138497897
15.9743102885699 0.184469484287
19.9498383822559 0.157397943633799
25.8898521722725 0.125649916180883
31.7947792366146 0.100623390080645
41.5366446596453 0.0680660519813239
51.1221755311639 0.0427725969692389
64.2107194517257 0.0150051916733168
};
\addplot [line width=1pt, color2, opacity=0.7, forget plot]
table {%
1.9999498009741 0.160803343747374
3.99959841686457 0.128414007741087
5.99864470794513 0.109471267552426
7.99678762519743 0.0960352299217948
9.99372625564787 0.0856176188433803
11.989159867684 0.0771100902464351
15.9743102885699 0.0636987007623483
19.9498383822559 0.0533127929023983
25.8898521722725 0.0411327662062763
31.7947792366146 0.0315314214781929
41.5366446596453 0.0190409054046349
51.1221755311639 0.00933715427422996
};
\node (c1) at (axis cs: 15,0.45) {$c \simeq 0.505$};
\coordinate (curve1) at (axis cs: 15,0.35);
\draw[->] (curve1) to[out = 20, in = -90] (c1);
\node (c2) at (axis cs: 90,0.30) {$c \simeq 0.487$};
\coordinate (curve2) at (axis cs: 15,0.22);
\draw[->] (curve2) to[out=40, in =180] (c2);
\node (ceff1) at (axis cs: 4,0.067) {$\tilde{c}_\mathrm{eff}\simeq 0.202$};
\node (ceff2) at (axis cs: 12,0.01) {$\tilde{c}_\mathrm{eff}\simeq 0.187$}; 
\coordinate (curve3) at (axis cs: 3,0.179);
\coordinate (curve4) at (axis cs: 15,0.058);
\draw[->] (curve3) to[out=-90] (ceff1); 
\draw[->] (curve4) to (ceff2);
\node (offset) at (axis cs: 6,0.55) {$ \simeq 0.125$};
\coordinate (offsetT) at (axis cs: 3, 0.495);
\coordinate (offsetB) at  (axis cs: 3, 0.40);
\coordinate (offsetM) at (axis cs: 3, 0.45);
\draw[<->] (offsetT) -- (offsetB);  
\draw (offsetM) to[out = 60, in = -90] (offset.-150);
\end{axis}
\end{tikzpicture}
    \end{minipage}
    \caption{\label{fig:EN}Scaling of the log-EN, $\mathcal{E}$, for infinite (left) and open (right) bcs. With increasing separation, $l$, the entanglement between $B_1$ and $B_2$ decreases logarithmically~[see Eq.~\eqref{eq:log_EN_eff}], as is expected from the intuition that correlation function of the degrees of freedom in $B_1$ and $B_2$ diminishes algebraically. The effective central charges, $\tilde{c}_{\rm eff}$, obtained from the scaling of the log-EN is different from that obtained from the scaling of the interface EE. The fit errors obtained are in the fourth decimal places. Note also that the offset for the duality defects are {\it smaller} than those for the energy defects of corresponding strengths. This is in contrast to the corresponding results for the interface EE~(see Fig.~\ref{fig:int_EE_scal} and maintext for further discussion). For the duality defect results with open bc, as for the interface EE, the presence of zero energy modes leads to deviations from the logarithmic scaling of the log-EN.  This leads to poorer estimates of the effective central charges for the open bc case compared to the infinite case. }
\end{figure}
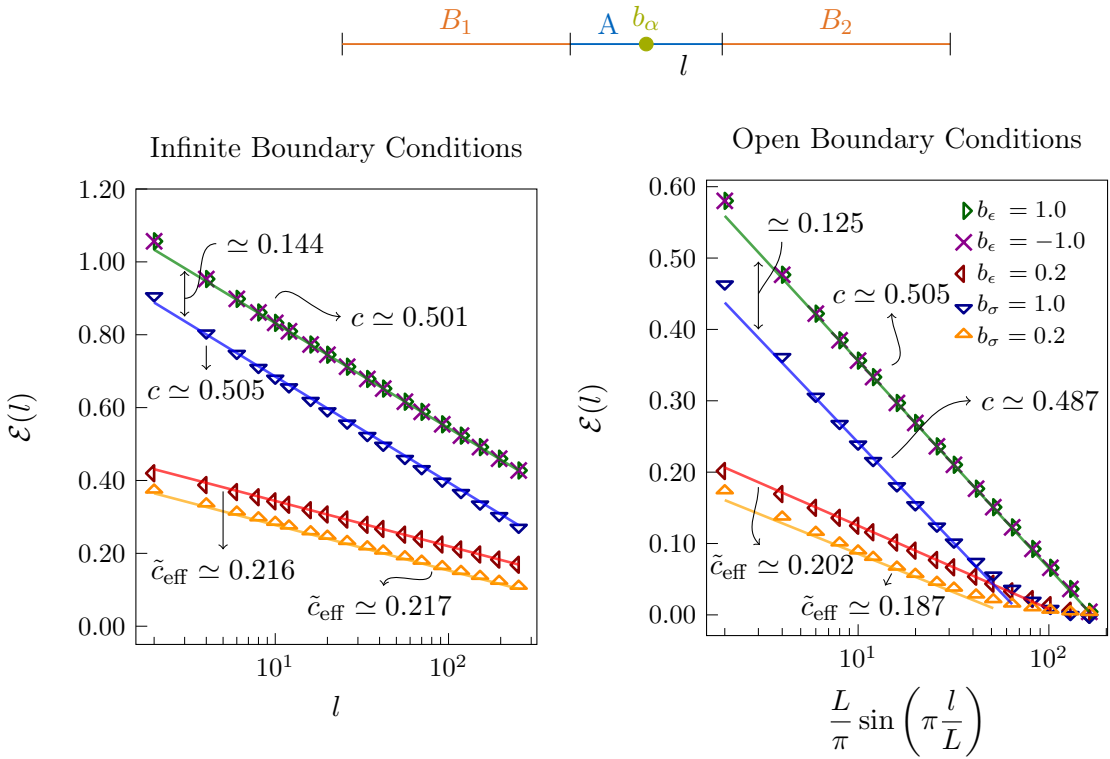

\begin{figure}
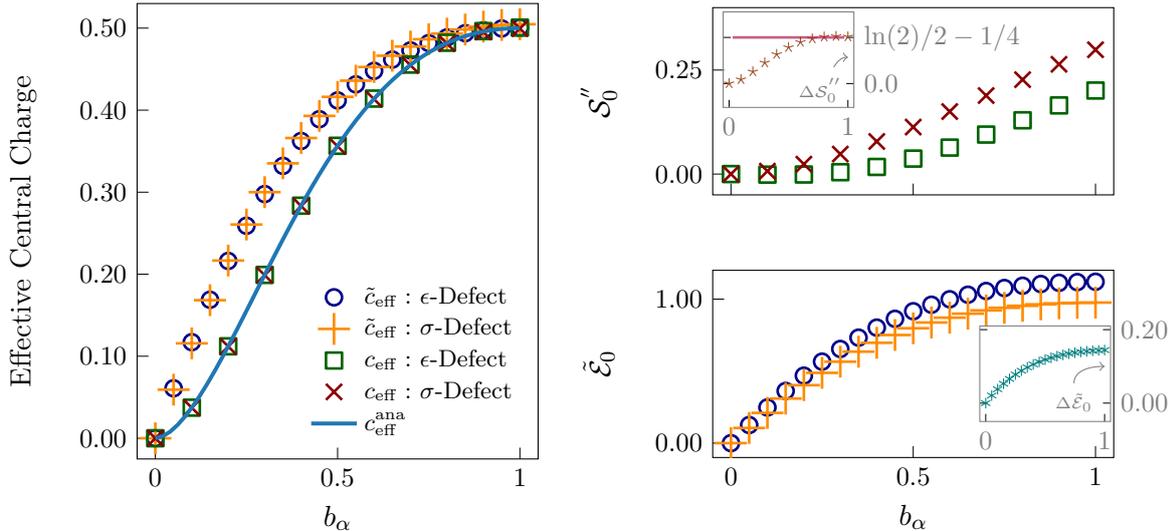

    \centering
    \hspace{-6em}
    \begin{minipage}{.45\linewidth}
\begin{tikzpicture}[
baseline=(current bounding box.center)
]

\definecolor{color0}{rgb}{1,0.549019607843137,0}
\definecolor{color1}{rgb}{0.12156862745098,0.466666666666667,0.705882352941177}

\begin{axis}[
height=3.0in,
legend cell align={left},
legend style={
  fill opacity=0.8,
  draw opacity=1,
  text opacity=1,
  at={(0.97,0.03)},
  anchor=south east,
  draw=none
},
minor xtick={},
minor ytick={},
tick pos=left,
width=2.7in,
x grid style={white!69.0196078431373!black},
xlabel={\(\displaystyle b_\alpha\)},
xmin=-0.05, xmax=1.05,
xtick style={color=black},
xtick={-0.5,0,0.5,1,1.5},
y grid style={white!69.0196078431373!black},
ylabel={\text{Effective Central Charge}},
ymin=-0.0252368490272275, ymax=0.529973829571665,
ytick style={color=black},
ytick={-0.1,0,0.1,0.2,0.3,0.4,0.5,0.6}
]
\addplot [draw=blue!54.5098039215686!black, mark=o, only marks]
table{%
x  y
1 0.500546761463852
0.95 0.499560846805399
0.9 0.497373060975105
0.85 0.493816458757386
0.8 0.488684207933954
0.75 0.481767114355913
0.7 0.472837822872568
0.65 0.461643816122047
0.6 0.447934505932316
0.55 0.43143677830489
0.5 0.411865542632163
0.45 0.388960311311171
0.4 0.362452213832913
0.35 0.332089660199163
0.3 0.297693541544081
0.25 0.259084201913707
0.2 0.216166801709464
0.15 0.168848495632415
0.1 0.117130762535725
0.05 0.060920268446242
0 -2.85892138107133e-15
};
\addlegendentry{$\tilde{c}_\mathrm{eff}:$ $\epsilon$-Defect}
\addplot [draw=color0, fill=color0, mark=+, only marks]
table{%
x  y
1 0.504736980544442
0.95 0.503869979497718
0.9 0.501797849665632
0.85 0.49834147836892
0.8 0.493291977612472
0.75 0.486440596098035
0.7 0.477542388904629
0.65 0.46634284837632
0.6 0.452580444221489
0.55 0.435951232286338
0.5 0.416179119113443
0.45 0.392963374150815
0.4 0.366038334059703
0.35 0.335139683568701
0.3 0.300023108257151
0.25 0.260561814601454
0.2 0.216652014395364
0.15 0.168315443408876
0.1 0.115719596860564
0.05 0.0592537993537276
0 -5.1121803000309e-15
};
\addlegendentry{$\tilde{c}_\mathrm{eff}:$ $\sigma$-Defect}
\addplot [draw=green!39.2156862745098!black, mark=square, only marks]
table{%
x  y
1 0.500598331915878
0.9 0.496369023530466
0.8 0.482194830097983
0.7 0.455532969100521
0.6 0.414091549323107
0.5 0.356536688700878
0.4 0.283547167354913
0.3 0.199146358430007
0.2 0.11209795104221
0.1 0.0372906169427225
0 0
};
\addlegendentry{$c_\mathrm{eff}:$ $\epsilon$-Defect}
\addplot [draw=red!54.5098039215686!black, fill=red!54.5098039215686!black, mark=x, only marks]
table{%
x  y
1 0.499530083557325
0.9 0.495334300298315
0.8 0.481218765740143
0.7 0.454643597869975
0.6 0.413316059222133
0.5 0.355896837266088
0.4 0.283055659003871
0.3 0.198806040743123
0.2 0.111891205146549
0.1 0.0371914696921808
0 0
};
\addlegendentry{$c_\mathrm{eff}:$ $\sigma$-Defect}
\addplot [line width=1.5pt, color1]
table {%
0 0
0.0204081632653061 0.00237814541511919
0.0408163265306122 0.00809682988085758
0.0612244897959184 0.0163323965198133
0.0816326530612245 0.0266219292544577
0.102040816326531 0.038622692072824
0.122448979591837 0.0520545520682597
0.142857142857143 0.0666765351183976
0.163265306122449 0.0822752320726981
0.183673469387755 0.0986584172913904
0.204081632653061 0.115651289028507
0.224489795918367 0.133094142441641
0.244897959183673 0.150840865530499
0.26530612244898 0.168757919702189
0.285714285714286 0.1867236062739
0.306122448979592 0.204627497598503
0.326530612244898 0.222369957062776
0.346938775510204 0.239861700402504
0.36734693877551 0.257023368926423
0.387755102040816 0.273785097269448
0.408163265306122 0.290086066418288
0.428571428571429 0.305874038275398
0.448979591836735 0.321104871756223
0.469387755102041 0.335742022869614
0.489795918367347 0.34975603276253
0.510204081632653 0.363124008564014
0.530612244897959 0.375829102219977
0.551020408163265 0.387859992505022
0.571428571428571 0.399210375135062
0.591836734693878 0.409878465466598
0.612244897959184 0.419866517719436
0.63265306122449 0.429180364049793
0.653061224489796 0.437828976169001
0.673469387755102 0.445824051579054
0.693877551020408 0.453179625901572
0.714285714285714 0.45991171222677
0.73469387755102 0.466037967913365
0.755102040816326 0.471577388834651
0.775510204081633 0.476550030692288
0.795918367346939 0.480976756707069
0.816326530612245 0.484879010742542
0.836734693877551 0.488278614719005
0.857142857142857 0.491197589027077
0.877551020408163 0.493657994546596
0.897959183673469 0.495681794812269
0.918367346938775 0.497290736836893
0.938775510204082 0.498506249100659
0.959183673469388 0.499349355236046
0.979591836734694 0.499840601977571
1 0.500000000001216
};
\addlegendentry{$c_{\mathrm{eff}}^\mathrm{ana}$}
\end{axis}

\end{tikzpicture}
    \end{minipage}
    \begin{minipage}{.45\linewidth}
        \vspace{1 pt}
        \hspace{14 pt}
\begin{tikzpicture}[
baseline=(current bounding box.center)
]

\definecolor{color0}{rgb}{1,0.549019607843137,0}

\begin{groupplot}[group style={group size=1 by 2}]
\nextgroupplot[
height=1.6in,
minor xtick={},
minor ytick={},
scaled x ticks=manual:{}{\pgfmathparse{#1}},
tick pos=left,
width=2.7in,
x grid style={white!69.0196078431373!black},
xmin=-0.05, xmax=1.05,
xtick style={color=black},
xtick={-0.5,0,0.5,1,1.5},
xticklabels={},
y grid style={white!69.0196078431373!black},
ylabel={\(\displaystyle \mathcal{S}^{''}_0\)},
ymin=-0.05, ymax=0.4,
ytick style={color=black},
ytick={-0.25,0,0.25,0.5}
]
\addplot [draw=green!39.2156862745098!black, mark=square, only marks]
table{%
x  y
1 0.200775481008024
0.9 0.165001050687156
0.8 0.129271780423753
0.7 0.0949274436071335
0.6 0.0636226222637844
0.5 0.0371582846388243
0.4 0.0171451384251838
0.3 0.00450949398827903
0.2 -0.00103727746980125
0.1 -0.00134298758617161
0 0
};
\addplot [draw=red!54.5098039215686!black, fill=red!54.5098039215686!black, mark=x, only marks]
table{%
x  y
1 0.298814417947244
0.9 0.263845665849222
0.8 0.226910926790823
0.7 0.188708468062233
0.6 0.150268499094323
0.5 0.112945951125537
0.4 0.0783359001926443
0.3 0.0481101699280301
0.2 0.0238604113130947
0.1 0.00708769739359891
0 0
};
\coordinate (insetPositionEntr) at (rel axis cs:0.0,0.98);

\nextgroupplot[
height=1.6in,
minor xtick={},
minor ytick={},
tick pos=left,
width=2.7in,
x grid style={white!69.0196078431373!black},
xlabel={\(\displaystyle b_\alpha\)},
xmin=-0.05, xmax=1.05,
xtick style={color=black},
xtick={-0.5,0,0.5,1,1.5},
y grid style={white!69.0196078431373!black},
ylabel={\(\displaystyle \tilde{\mathcal{E}}_0\)},
ymin=-0.1, ymax=1.2,
ytick style={color=black},
ytick={-1,0,1,2}
]
\addplot [draw=blue!54.5098039215686!black, mark=o, only marks]
table{%
x  y
1 1.12021286094795
0.95 1.11786709618754
0.9 1.11277360274376
0.85 1.10453886248546
0.8 1.09268769765198
0.75 1.07674618183418
0.7 1.05616704083072
0.65 1.03037435052586
0.6 0.998790649351775
0.55 0.960776041581364
0.5 0.91570312999764
0.45 0.862977477806965
0.4 0.801987900011746
0.35 0.732316240132068
0.3 0.653598973548682
0.25 0.565632196449289
0.2 0.468368543792685
0.15 0.362173008212944
0.1 0.247636194963599
0.05 0.12576376835618
0 -3.37416869147426e-15
};
\addplot [draw=color0, fill=color0, mark=+, only marks]
table{%
x  y
1 0.975886615708188
0.95 0.974349746401821
0.9 0.97034288251368
0.85 0.963486750637899
0.8 0.95336837372146
0.75 0.93955290723809
0.7 0.92156310783524
0.65 0.898884032507098
0.6 0.870989723647747
0.55 0.8373433681048
0.5 0.797404553545526
0.45 0.750651104719914
0.4 0.696645608687357
0.35 0.635009585147028
0.3 0.565428722945897
0.25 0.487869581458274
0.2 0.402392337970264
0.15 0.309431124628187
0.1 0.20982746131055
0.05 0.105090633945225
0 -3.04700030507449e-15
};
\coordinate (insetPositionNeg) at (rel axis cs:0.64,0.78);
\end{groupplot}
\input{22-02-17_offset_inlet_entr.tikz}
\input{22-02-17_offset_inlet_neg.tikz}
\end{tikzpicture}
    \end{minipage}
    \caption{\label{ceff_var} Variation of the effective central charges (left) and the sub-leading corrections (right) with the defect strength for the energy and duality defects. The effective central charges, $c_\mathrm{eff}$, obtained from the scaling of the interface EE, $\mathcal{S}_\mathrm{I}$,  for the energy and duality defects are identical for a given defect strength and are compatible with the predictions of Eq. \eqref{ceff}, denoted in this plot by $c_{\rm eff}^{\rm ana}$. The computed values $\tilde{c}_\mathrm{eff}$ from the log-EN again agree in between the defects of equal strength. The subleading corrections on the right increase with their difference (shown in the respective inserts) eventually saturating towards a fixed value. For the interface EE, this offset $\Delta \mathcal{S}^{''}_0$ is in good agreement with the analytic prediction for $b_\sigma=1$ of Eq.~\eqref{eq:DeltaInterfaceEntr}, shown as the pink line. The corresponding subleading corrections for the log-EN are shown in the bottom right panel.}
\end{figure}

The log-EN is computed from the ground state obtained using DMRG~(see Appendix~\ref{sec:App_DMRG} for technical details). Fig.~\ref{fig:EN} shows the results for the energy and the duality defects. As expected, the results are identical for $b_\epsilon=\pm1$~(see Sec.~\ref{sec:DMRG_res} for similar results for the EE), with $\tilde{c}_{\rm eff}=1/2$~(see Fig.~\ref{ceff_var}). Lowering the defect strength leads to a monotonically decreasing $\tilde{c}_{\rm eff}$ culminating in $\tilde{c}_{\rm eff}=0$ for zero defect strength, when the 1D system is cut into half and there is zero entanglement between $B_1$ and $B_2$. As is evident from Fig.~\ref{ceff_var}, $\tilde{c}_{\rm eff}\geq c_{\rm eff}$ with equality being satisfied only for perfectly reflective and perfectly transmissive defects. When the size of $A\cup B_1\cup B_2$ is finite~(say $L$), the scaling is similar after the substitution of $l$ by $L\sin(\pi l/L)/\pi$. 

The duality defect for the corresponding defect strengths gives rise to identical $\tilde{c}_{\rm eff}$ as the energy defect. Recall that the same was also observed for $c_{\rm eff}$ during the computation of the interface EE~(see also Fig.~\ref{ceff_var}). This can be viewed as a consequence of the fact that the microscopic Hamiltonians for the two kinds of defects are essentially identical barring the zero energy modes for the duality defect. This is clearly visible in the fermionic formulation~(see Sec.~\ref{sec:defects}). However, as for the EE, the zero energy modes give rise to two distinct effects. First, they affect the subleading term, $\tilde{\cal E}_0$ in Eq.~\eqref{eq:log_EN}. This manifests itself in a difference between the offsets for energy and duality defects for a given defect strength~[see Fig.~\ref{fig:EN}, infinite bc and $l\ll L$ part of the open bc]. Interestingly, the subleading term is {\it lower} for the duality defect compared to the energy defect for the same defect strength. This is in contrast to the symmetric and interface EEs, where the subleading term was higher for the duality defect. In fact, the higher symmetric EE due to the unpaired zero mode within the interval A is directly related to the lower log-EN between $B_1$ and $B_2$. Intuitively, the enhanced EE of A must be accompanied by a lowering of entanglement between $B_1$ and $B_2$ since the system overall remains in a pure state. Second, the local and nonlocal zero modes of the duality defect give rise to finite-size corrections to the log-EN, similar to what is described for the EE in Sec.~\ref{sec:DMRG_res}. This is visible in the deviations from the logarithmic scaling the log-EN for the open bc~(Fig.~\ref{fig:EN}, right panel, $l\sim L$). Notice there is no such effect for the energy defect with open bc since the latter has no zero modes. Fig.~\ref{ceff_var} summarizes the results for the effective central charges and offsets computed for the interface EE and log-EN. 

\section{Conclusion and Outlook}
\label{sec:concl} 
To summarize, we analyzed the rich entanglement properties of the Ising CFT in the presence of defects. In particular, we computed the EEs~(symmetric and interface) and the log-EN for subsystems in the presence of energy and duality defects using DMRG. We showed that for the duality defect, due to the existence of zero energy modes, the EEs and the log-EN respectively receive positive and negative additional contributions compared to the energy defect of the same defect strength. For finite systems, the same zero energy modes also give rise to finite-size corrections for both the EE and the log-EN, which lead to deviations from the usual logarithmic scaling that is characteristic of one-dimensional critical systems. 

Before concluding, we present some potential future directions of research. First, we computed the log-EN numerically using DMRG for the case with defects. In the absence of defects, the log-EN can be computed using replica techniques for CFTs~\cite{Calabrese2012, Calabrese2013_EN, Calabrese:2013mi}. Potentially, the existing computations of the interface EE~\cite{Sakai2008, Brehm2015} can be generalized to compute the leading term of the logarithmic scaling of the log-EN. The computations of the subleading term for the log-EN and the associated finite-size corrections due to the zero energy modes remain also open problems, which are likely beyond the scope of the aforementioned computation. Second, in this work, we have focussed on the EE and the log-EN as entanglement measures for the defect setting. Yet another quantity of particular significance is the entanglement or modular Hamiltonian~\cite{Haag2012}. The latter plays a central role in the analysis of topological phases of matter in 2+1 space-time dimensions~\cite{Li2008}. For 1+1D CFTs, the spectra of the entanglement Hamiltonians can be related to the Hamiltonian spectra of corresponding boundary CFTs~\cite{Cardy2016, Alba2017, Roy2020a}. It will be interesting to generalize these results to the case of defects. We note that some progress has been made for massless Dirac fermions~\cite{Mintchev:2020jhc} and the chiral fermions with zero modes~(equivalently the antiperiodic Ising chain)~\cite{Klich2017}. Third, in recent works, the concept of entanglement Hamiltonian, which is typically written for the total system being in pure state, has been generalized to the case when the total system is mixed. This leads to the concept of negativity Hamiltonian~\cite{Ruggiero2016, Murciano:2022vhe}. It is an interesting open problem to consider generalizations of these concepts to the cases with defects. Fourth, we note that defect Hamiltonians for other rational CFTs contain more complicated set of zero modes~\cite{Belletete2020}. The question of subleading corrections in the EEs for these models remains open. Finally, the zero modes in the Ising model with the duality defects are reminiscent of the zero modes of gapless symmetry-protected-topological phases~\cite{Verresen2018}. It will be interesting to further analyze this relationship.We hope to return to some of these questions in the future. \\
\\
{\it Note added in proof}: After the completion of the paper, the authors became aware of the fact that the values of $\tilde{c}_{\rm eff}$ shown in Fig.~\ref{ceff_var} are in excellent agreement with the results of Refs.~\cite{peschelExactResultsEntanglement2012,gruberTimeEvolutionEntanglement2020}. The latter works analyze a different geometric configuration, that of two segments connected through a conformal defect. We are grateful to Viktor Eisler for bringing this to our attention.  
\acknowledgments
AR acknowledges discussions with Hubert Saleur and Johannes Hauschild. AR was supported from a grant from the Simons Foundation (825876, TDN). FP acknowledges support of the European Research Council (ERC) under the European Unions Horizon 2020 research and innovation program (grant agreement No. 771537). FP also acknowledges the support of the Deutsche Forschungsgemeinschaft (DFG, German Research Foundation) under Germany’s Excellence Strategy EXC-2111-390814868. FP's research is part of the Munich Quantum Valley, which is supported by the Bavarian state government with funds from the Hightech Agenda Bayern Plus.
\appendix
\section{Computation of Symmetric Entanglement Entropy and Negativity with Matrix Product States}
Density Matrix Renormalization Group (DMRG) calculations use the fact that ground states of a local 1D lattice Hamiltonian can be efficiently represented by Matrix Product States (MPS).
Such a state $\ket{\Psi}$ is nicely visualized in Penrose tensor notation as
\begin{align}
    \ket{\Psi} &= 
\begin{tikzpicture}[inner sep=1mm,baseline=(current  bounding  box.center)]
    \node (A 0) at (0, 0) {};
    \node (A 7) at (7, 0) {};
    \foreach \i in {1,...,6} {
        \node[tensorE] (A \i) at (\i, 0) {$A_{\i}$};
        \node (S \i) at (\i, -0.7) {};
        \draw[-] (A \i) -- (S \i);
    };
    \foreach \i in {1,...,5} {
        \pgfmathtruncatemacro{\iplusone}{\i + 1};
        \draw[-] (A \i) -- (A \iplusone);
    };
    \draw[dotted, line width = 0.5mm] (A 0) -- (A 1);
    \draw[dotted, line width = 0.5mm] (A 6) -- (A 7);
\end{tikzpicture}.
\end{align}
Each box is a tensor of rank equal to the number of legs connected to it. In this example, all tensors are of rank 3.
The unconnected lose legs represent the physical dimension $D$ which is in the case of Spin-$\frac{1}{2}$ Systems $D=2$.
The horizontal virtual bonds are of dimension $\chi$, the bond dimension.
The bond dimension is a controllable parameter directly governing the entanglement captured by the MPS.\\
In the following we assume that $\ket{\Psi}$ is in left canonical form such that
\begin{align}
    \begin{tikzpicture}[inner sep=1mm,baseline=(current  bounding  box.center)]
    \node[tensorE] (A 1) at (0, 0) {$A_{i}$};
    \coordinate (aL) at (-0.7,-0.5);
    \node[tensorE] (Adag 1) at (0,-1) {$A_{i}^{\ast}$};
    \coordinate (A 2) at (0.7, 0);
    \coordinate (Adag 2) at (0.7,-1);
    \draw[-] (A 1) -- (A 2);
    \draw[-] (Adag 1) -- (Adag 2);
    \draw[-] (A 1) -- (Adag 1);
    \draw (A 1.180) to [out=180, in=90](aL);
    \draw (aL) to [out=-90, in=180](Adag 1); 
\end{tikzpicture} = \;\begin{tikzpicture}[inner sep=1mm,baseline=(current  bounding  box.center)]
    \coordinate  (A 1) at (0, 0) {};
    \node[tensorE] (aL) at (-0.7,-0.5) {$\mathbb{1}$};
    \coordinate  (Adag 1) at (0,-1) {};
    \draw (A 1) to [out=180, in=90](aL.north);
    \draw (Adag 1) to [out=180,in=-90](aL.south); 
\end{tikzpicture}
\end{align}
and 
\begin{align}
    \begin{tikzpicture}[inner sep=1mm,baseline=(current  bounding  box.center)]
    \node[tensorE] (A 1) at (0, 0) {$A_{i}$};
    \coordinate  (aR) at (0.7,-0.5) {};
    \node[tensorE] (Adag 1) at (0,-1) {$A_{i}^{\ast}$};
    \coordinate  (A 2) at (-0.7, 0) {};
    \coordinate  (Adag 2) at (-0.7,-1) {};
    \draw[-] (A 1) -- (A 2);
    \draw[-] (Adag 1) -- (Adag 2);
    \draw[-] (A 1) -- (Adag 1);
    \draw (A 1.0) to [out=0, in=90](aR.90) to [out=-90, in=0](Adag 1); 
\end{tikzpicture} = \begin{tikzpicture}[inner sep=1mm,baseline=(current  bounding  box.center)]
    \coordinate (A 1) at (0, 0) {};
    \node[tensorE] (aR) at (0.7,-0.5) {$\Lambda^2$};
    \coordinate  (Adag 1) at (0,-1) {};
    \draw (A 1) to [out=0, in=90](aR.north);
    \draw (Adag 1) to [out=0, in=-90](aR.south); 
\end{tikzpicture}.   
\end{align}
$\Lambda$ is a diagonal matrix carrying the Schmidt values of a bipartition on that bond.
This efficient tensor network representation can be used to construct reduced density operators in an economical way \cite{ruggieroEntanglementNegativityRandom2016, wichterichScalingEntanglementSeparated}.
Using the formal definition of the density operator
\begin{align}
    \hat{\rho} &= \ketbra{\Psi}{\Psi}
\end{align}
for matrix product states gives
\begin{align}
    \hat{\rho} &= 
\begin{tikzpicture}[inner sep=1mm,baseline=(current  bounding  box.center)]
    \node (Adag 0) at (0, 1) {};
    \node (Adag 7) at (7, 1) {};
    \foreach \i in {1,...,6} {
        \node[tensorE] (Adag \i) at (\i, 1) {$A^{\ast}_{\i}$};
        \node (Sdag \i) at (\i, 1.7) {};
        \draw[-] (Adag \i) -- (Sdag \i);
    };
    \foreach \i in {1,...,5} {
        \pgfmathtruncatemacro{\iplusone}{\i + 1};
        \draw[-] (Adag \i) -- (Adag \iplusone);
    };
    \draw[dotted, line width = 0.5mm] (Adag 0) -- (Adag 1);
    \draw[dotted, line width = 0.5mm] (Adag 6) -- (Adag 7);

    \node (A 0) at (0, 0) {};
    \node (A 7) at (7, 0) {};
    \foreach \i in {1,...,6} {
        \node[tensorE] (A \i) at (\i, 0) {$A_{\i}$};
        \node (S \i) at (\i, -0.7) {};
        \draw[-] (A \i) -- (S \i);
    };
    \foreach \i in {1,...,5} {
        \pgfmathtruncatemacro{\iplusone}{\i + 1};
        \draw[-] (A \i) -- (A \iplusone);
    };
    \draw[dotted, line width = 0.5mm] (A 0) -- (A 1);
    \draw[dotted, line width = 0.5mm] (A 6) -- (A 7);
\end{tikzpicture}
\end{align}
the so-called matrix product density operator.
As each tensor represent a specific lattice site, different system partitions can be applied to the operator.\\
The Symmetric Entanglement entropy is obtained by assigning a central segment to subsystem A (red) and the rest to subsystem B (blue):
\begin{align}
    \hat{\rho} &= \begin{tikzpicture}[inner sep=1mm,baseline=(current  bounding  box.center)]
    \node (Adag 0) at (0, 1) {};
    \node (Adag 7) at (7, 1) {};
    \foreach \i in {1,...,2} {
        \node[tensorB] (Adag \i) at (\i, 1) {$A^{\ast}_{\i}$};
        \node (Sdag \i) at (\i, 1.7) {};
        \draw[-] (Adag \i) -- (Sdag \i);
    };
    \foreach \i in {3,...,4} {
        \node[tensorA] (Adag \i) at (\i, 1) {$A^{\ast}_{\i}$};
        \node (Sdag \i) at (\i, 1.7) {};
        \draw[-] (Adag \i) -- (Sdag \i);
    };
    \foreach \i in {5,...,6} {
        \node[tensorB] (Adag \i) at (\i, 1) {$A^{\ast}_{\i}$};
        \node (Sdag \i) at (\i, 1.7) {};
        \draw[-] (Adag \i) -- (Sdag \i);
    };
    \foreach \i in {1,...,5} {
        \pgfmathtruncatemacro{\iplusone}{\i + 1};
        \draw[-] (Adag \i) -- (Adag \iplusone);
    };
    \draw[dotted, line width = 0.5mm] (Adag 0) -- (Adag 1);
    \draw[dotted, line width = 0.5mm] (Adag 6) -- (Adag 7);

    \node (A 0) at (0, 0) {};
    \node (A 7) at (7, 0) {};
    \foreach \i in {1,...,2} {
        \node[tensorB] (A \i) at (\i, 0) {$A_{\i}$};
        \node (S \i) at (\i, -0.7) {};
        \draw[-] (A \i) -- (S \i);
    };
    \foreach \i in {3,...,4} {
        \node[tensorA] (A \i) at (\i, 0) {$A_{\i}$};
        \node (S \i) at (\i, -0.7) {};
        \draw[-] (A \i) -- (S \i);
    };
    \foreach \i in {5,...,6} {
        \node[tensorB] (A \i) at (\i, 0) {$A_{\i}$};
        \node (S \i) at (\i, -0.7) {};
        \draw[-] (A \i) -- (S \i);
    };
    \foreach \i in {1,...,5} {
        \pgfmathtruncatemacro{\iplusone}{\i + 1};
        \draw[-] (A \i) -- (A \iplusone);
    };
    \draw[dotted, line width = 0.5mm] (A 0) -- (A 1);
    \draw[dotted, line width = 0.5mm] (A 6) -- (A 7);

    \draw[dashed] (2.5,1.7) -- (2.5,-0.7);
    \draw[dashed] (4.5,1.7) -- (4.5,-0.7);
\end{tikzpicture} .
\end{align}
After tracing over the subsystem A
\begin{align}
    \Trace_A{\hat{\rho}} &= \begin{tikzpicture}[inner sep=1mm,baseline=(current  bounding  box.center)]
    \node (Adag 0) at (0, 1) {};
    \node (Adag 7) at (7, 1) {};
    \foreach \i in {1,...,2} {
        \node[tensorB] (Adag \i) at (\i, 1) {$A^{\ast}_{\i}$};
        \node (Sdag \i) at (\i, 1.7) {};
        \draw[-] (Adag \i) -- (Sdag \i);
    };
    \foreach \i in {3,...,4} {
        \node[tensorA] (Adag \i) at (\i, 1) {$A^{\ast}_{\i}$};
    };
    \foreach \i in {5,...,6} {
        \node[tensorB] (Adag \i) at (\i, 1) {$A^{\ast}_{\i}$};
        \node (Sdag \i) at (\i, 1.7) {};
        \draw[-] (Adag \i) -- (Sdag \i);
    };
    \foreach \i in {1,...,5} {
        \pgfmathtruncatemacro{\iplusone}{\i + 1};
        \draw[-] (Adag \i) -- (Adag \iplusone);
    };
    \draw[dotted, line width = 0.5mm] (Adag 0) -- (Adag 1);
    \draw[dotted, line width = 0.5mm] (Adag 6) -- (Adag 7);

    \node (A 0) at (0, 0) {};
    \node (A 7) at (7, 0) {};
    \foreach \i in {1,...,2} {
        \node[tensorB] (A \i) at (\i, 0) {$A_{\i}$};
        \node (S \i) at (\i, -0.7) {};
        \draw[-] (A \i) -- (S \i);
    };
    \foreach \i in {3,...,4} {
        \node[tensorA] (A \i) at (\i, 0) {$A_{\i}$};
    };
    \foreach \i in {5,...,6} {
        \node[tensorB] (A \i) at (\i, 0) {$A_{\i}$};
        \node (S \i) at (\i, -0.7) {};
        \draw[-] (A \i) -- (S \i);
    };
    \foreach \i in {1,...,5} {
        \pgfmathtruncatemacro{\iplusone}{\i + 1};
        \draw[-] (A \i) -- (A \iplusone);
    };
    \foreach \i in {3,..., 4}{
        \draw[-] (A \i) -- (Adag \i);
    };
    \draw[dotted, line width = 0.5mm] (A 0) -- (A 1);
    \draw[dotted, line width = 0.5mm] (A 6) -- (A 7);

    \draw[dashed] (2.5,1.7) -- (2.5,-0.7);
    \draw[dashed] (4.5,1.7) -- (4.5,-0.7);
\end{tikzpicture}
\end{align}
the center can be represented very similar to a transfer matrix $T_A$
\begin{align}
    \Trace_A{\hat{\rho}} &= \begin{tikzpicture}[inner sep=1mm,baseline=(current  bounding  box.center)]
    \node (Adag 0) at (0, 1) {};
    \node (Adag 7) at (7, 1) {};
    \node[transferA] (T) at (3.5, 0.5) {$T_A$};
    \foreach \i in {1,...,2} {
        \node[tensorB] (Adag \i) at (\i, 1) {$A^{\ast}_{\i}$};
        \node (Sdag \i) at (\i, 1.7) {};
        \draw[-] (Adag \i) -- (Sdag \i);
    };
    \foreach \i in {5,...,6} {
        \node[tensorB] (Adag \i) at (\i, 1) {$A^{\ast}_{\i}$};
        \node (Sdag \i) at (\i, 1.7) {};
        \draw[-] (Adag \i) -- (Sdag \i);
    };
    \foreach \i in {1,...,1} {
        \pgfmathtruncatemacro{\iplusone}{\i + 1};
        \draw[-] (Adag \i) -- (Adag \iplusone);
    };
    \foreach \i in {5,...,5} {
        \pgfmathtruncatemacro{\iplusone}{\i + 1};
        \draw[-] (Adag \i) -- (Adag \iplusone);
    };
    \draw[dotted, line width = 0.5mm] (Adag 0) -- (Adag 1);
    \draw[dotted, line width = 0.5mm] (Adag 6) -- (Adag 7);

    \node (A 0) at (0, 0) {};
    \node (A 7) at (7, 0) {};
    \foreach \i in {1,...,2} {
        \node[tensorB] (A \i) at (\i, 0) {$A_{\i}$};
        \node (S \i) at (\i, -0.7) {};
        \draw[-] (A \i) -- (S \i);
    };
    \foreach \i in {5,...,6} {
        \node[tensorB] (A \i) at (\i, 0) {$A_{\i}$};
        \node (S \i) at (\i, -0.7) {};
        \draw[-] (A \i) -- (S \i);
    };
    \foreach \i in {1,...,1} {
        \pgfmathtruncatemacro{\iplusone}{\i + 1};
        \draw[-] (A \i) -- (A \iplusone);
    };
    \foreach \i in {5,...,5} {
        \pgfmathtruncatemacro{\iplusone}{\i + 1};
        \draw[-] (A \i) -- (A \iplusone);
    };
    \draw[dotted, line width = 0.5mm] (A 0) -- (A 1);
    \draw[dotted, line width = 0.5mm] (A 6) -- (A 7);

    \draw[dashed] (2.5,1.7) -- (2.5,-0.7);
    \draw[dashed] (4.5,1.7) -- (4.5,-0.7);

    \draw[-] (A 2) -- (T);
    \draw[-] (Adag 2) -- (T);
    \draw[-] (A 5) -- (T);
    \draw[-] (Adag 5) -- (T);
\end{tikzpicture}.
\end{align}
The possibly infinite segments of subsystem B can be compressed without any losses by using the properties of the canonical form. This can be understood by first tracing over each half of B as well:
\begin{align}
    \tilde{\Trace}_A{\hat{\rho}} =  \begin{tikzpicture}[inner sep=1mm,baseline=(current  bounding  box.center)]
    \node (Adag 0) at (-1, 1) {};
    \node (Adag 7) at (8, 1) {};
    \node (A 0) at (-1, 0) {};
    \node (A 7) at (8, 0) {};
    \node[transferA] (T) at (3.5, 0.5) {$T_A$};
    \foreach \i in {1,...,2} {
        \pgfmathtruncatemacro{\iminusone}{\i - 1};
        \node[tensorB] (Adag \i) at (\iminusone, 1) {$A^{\ast}_{\i}$};
        \node[tensorB] (A \i) at (\iminusone, 0) {$A_{\i}$};
        \draw[-] (Adag \i) -- (A \i);
    };
    \foreach \i in {5,...,6} {
        \pgfmathtruncatemacro{\iplusone}{\i + 1};
        \node[tensorB] (Adag \i) at (\iplusone, 1) {$A^{\ast}_{\i}$};
        \node[tensorB] (A \i) at (\iplusone, 0) {$A_{\i}$};
        \draw[-] (Adag \i) -- (A \i);
    };
    \foreach \i in {1,...,1} {
        \pgfmathtruncatemacro{\iplusone}{\i + 1};
        \draw[-] (Adag \i) -- (Adag \iplusone);
        \draw[-] (A \i) -- (A \iplusone);
    };
    \foreach \i in {5,...,5} {
        \pgfmathtruncatemacro{\iplusone}{\i + 1};
        \draw[-] (Adag \i) -- (Adag \iplusone);
    };
    \draw[dotted, line width = 0.5mm] (Adag 0) -- (Adag 1);
    \draw[dotted, line width = 0.5mm] (Adag 6) -- (Adag 7);
    \draw[dotted, line width = 0.5mm] (A 0) -- (A 1);
    \draw[dotted, line width = 0.5mm] (A 6) -- (A 7);

    \node (leftGapT) at (2,1) {};
    \node  (leftGapB) at (2,0) {};
    \node (rightGapT) at (5,1) {};
    \node (rightGapB) at (5,0) {};
    \draw[dashed] (2,1.7) -- (2, -0.7);
    \draw[dashed] (5,1.7) -- (5,-0.7);

    \draw[-] (Adag 2) -- (leftGapT);
    \draw[-] (A 2) -- (leftGapB);
    \draw[-] (A 5) -- (A 6);

    \draw[-] (Adag 5) -- (rightGapT);
    \draw[-] (A 5) -- (rightGapB);

    \draw[-] (T) -- (leftGapT);
    \draw[-] (T) -- (leftGapB);
    \draw[-] (T) -- (rightGapT);
    \draw[-] (T) -- (rightGapB);

\end{tikzpicture}
\end{align}
Due to the canonical form this simplifies to
\begin{align}
    \tilde{\Trace}_A{\hat{\rho}} =  \begin{tikzpicture}[inner sep=1mm,baseline=(current  bounding  box.center)]

    \node[transferA] (T) at (3.5, 0.5) {$T_A$};
    \node[tensorB] (left) at (1,0.5) {$\mathbb{1}$};
    \node[tensorB] (right) at(6,0.5) {$\Lambda^2$};

    \node (leftGapT) at (2,1) {};
    \node  (leftGapB) at (2,0) {};
    \node (rightGapT) at (5,1) {};
    \node (rightGapB) at (5,0) {};
    \draw[dashed] (2,1.7) -- (2, -0.7);
    \draw[dashed] (5,1.7) -- (5,-0.7);

    \draw[-] (left) to[out=90, in = 180] (leftGapT);
    \draw[-] (left) to[out=-90, in = 180] (leftGapB);

    \draw[-] (right) to[out=90, in = 0] (rightGapT);
    \draw[-] (right) to[out=-90, in = 0] (rightGapB);

    \draw[-] (T) -- (leftGapT);
    \draw[-] (T) -- (leftGapB);
    \draw[-] (T) -- (rightGapT);
    \draw[-] (T) -- (rightGapB);

\end{tikzpicture}.
\end{align}
After splitting the outside matrices by matrix factorizing the compressed form of the reduced density matrix is
\begin{align}
    \Trace_A{\hat{\rho}} &= \begin{tikzpicture}[inner sep=1mm, baseline=(current  bounding  box.center)]
    \node[tensorB] (U 2) at (3, 1) {$\mathbb{1}$};
    \node[tensorB] (V 2) at (3, -1) {$\mathbb{1}$};
    \node[transferA] (T 3) at (5, 0) {$T_A$};
    \node[tensorB] (U 4) at (7, 1) {$\Lambda$};
    \node[tensorB] (V 4) at (7, -1) {$\Lambda$};
    \node (S 2) at (3,-2) {$\chi$};
    \node (Sdag 2) at (3,2) {$\chi$};
    \node (S 4) at (7,-2) {$\chi$};
    \node (Sdag 4) at (7,2) {$\chi$};

    \draw[-] (U 2) -- (Sdag 2);
    \draw[-] (V 2) -- (S 2);
    \draw[-] (U 4) -- (Sdag 4);
    \draw[-] (V 4) -- (S 4);
    \draw (T 3.north west) to [out=135, in=-60](U 2.-60);
    \draw (T 3.south west) to [out=-135, in=60](V 2.60);
    \draw (T 3.north east) to [out=45, in=-120](U 4.-120);
    \draw (T 3.south east) to [out=-45, in=120](V 4.120);
\end{tikzpicture} .
\end{align}
As DMRG already gives access to the Schmidt values $\Lambda$ at each bond, only the contraction of $T_A$ must be performed. This scales with $O(\chi^5)$ in computational complexity and $O(\chi^4)$ in memory.
The final operator is of dimension $\chi^4$ independent of the number of lattice sites involved. This is in contrast to a naive construction.
If the tensor network structure of the state is neglected the matrix will scale exponentially in memory with $D^{2l}$ where $l$ is the segment size. 
We want to emphasize that no additional approximations apart from the finite bond dimensions of the DMRG calculations were performed.  
A nearly identical method can be used to calculate the negativity of two segments touching the possibly infinite boundaries separated by a finite interval.
The segmentation used in this case is 
\begin{align}
    \hat{\rho} &= \begin{tikzpicture}[inner sep=1mm,baseline=(current  bounding  box.center)]
    \node (Adag 0) at (0, 1) {};
    \node (Adag 7) at (7, 1) {};
    \foreach \i in {1,...,2} {
        \node[tensorA] (Adag \i) at (\i, 1) {$A^{\ast}_{\i}$};
        \node (Sdag \i) at (\i, 1.7) {};
        \draw[-] (Adag \i) -- (Sdag \i);
    };
    \foreach \i in {3,...,4} {
        \node[tensorE] (Adag \i) at (\i, 1) {$A^{\ast}_{\i}$};
        \node (Sdag \i) at (\i, 1.7) {};
        \draw[-] (Adag \i) -- (Sdag \i);
    };
    \foreach \i in {5,...,6} {
        \node[tensorB] (Adag \i) at (\i, 1) {$A^{\ast}_{\i}$};
        \node (Sdag \i) at (\i, 1.7) {};
        \draw[-] (Adag \i) -- (Sdag \i);
    };
    \foreach \i in {1,...,5} {
        \pgfmathtruncatemacro{\iplusone}{\i + 1};
        \draw[-] (Adag \i) -- (Adag \iplusone);
    };
    \draw[dotted, line width = 0.5mm] (Adag 0) -- (Adag 1);
    \draw[dotted, line width = 0.5mm] (Adag 6) -- (Adag 7);

    \node (A 0) at (0, 0) {};
    \node (A 7) at (7, 0) {};
    \foreach \i in {1,...,2} {
        \node[tensorA] (A \i) at (\i, 0) {$A_{\i}$};
        \node (S \i) at (\i, -0.7) {};
        \draw[-] (A \i) -- (S \i);
    };
    \foreach \i in {3,...,4} {
        \node[tensorE] (A \i) at (\i, 0) {$A_{\i}$};
        \node (S \i) at (\i, -0.7) {};
        \draw[-] (A \i) -- (S \i);
    };
    \foreach \i in {5,...,6} {
        \node[tensorB] (A \i) at (\i, 0) {$A_{\i}$};
        \node (S \i) at (\i, -0.7) {};
        \draw[-] (A \i) -- (S \i);
    };
    \foreach \i in {1,...,5} {
        \pgfmathtruncatemacro{\iplusone}{\i + 1};
        \draw[-] (A \i) -- (A \iplusone);
    };
    \draw[dotted, line width = 0.5mm] (A 0) -- (A 1);
    \draw[dotted, line width = 0.5mm] (A 6) -- (A 7);

    \draw[dashed] (2.5,1.7) -- (2.5,-0.7);
    \draw[dashed] (4.5,1.7) -- (4.5,-0.7);
\end{tikzpicture}
\end{align}
where segment $B_1$ is shown in red, segment $B_2$ in blue and the separating gap $A$ in gray. After following analogous steps to the above, this results in
\begin{align}
    \Trace_A{\hat{\rho}}^{T_{B_2}} &= \begin{tikzpicture}[inner sep=1mm,baseline=(current  bounding  box.center)]
    \node[tensorA] (U 2) at (3, 1) {$\mathbb{1}$};
    \node[tensorA] (V 2) at (3, -1) {$\mathbb{1}$};
    \node[transferE] (T 3) at (5, 0) {$T_A$};
    \node[tensorB] (V 4) at (7, 1) {$\Lambda$};
    \node[tensorB] (U 4) at (7, -1) {$\Lambda$};
    \node (S 2) at (3,-2) {$\chi$};
    \node (Sdag 2) at (3,2) {$\chi$};
    \node (S 4) at (7,-2) {$\chi$};
    \node (Sdag 4) at (7,2) {$\chi$};

    \draw[-] (U 2) -- (Sdag 2);
    \draw[-] (V 2) -- (S 2);
    \draw[-] (U 4) -- (S 4);
    \draw[-] (V 4) -- (Sdag 4);

    \draw (T 3.north west) to [out=135, in=-60](U 2.-60);
    \draw (T 3.south west) to [out=-135, in=60](V 2.60);
    \draw (T 3.north east) to [out=45, in=120](U 4.120);
    \draw (T 3.south east) to [out=-45, in=-120](V 4.-120);
\end{tikzpicture}.
\end{align}
After tracing over the separating interval the individual subsystems are still individually accessible. This allows to partially transpose segment $B_2$ by twisting the corresponding legs.
\label{sec:App_DMRG}

\bibliographystyle{JHEP}
\bibliography{library_1}

\end{document}